\documentclass[english,aps,pra,preprint,groupedaddress,showpacs,showkeys]{revtex4-1}
\usepackage[latin9]{inputenc}
\usepackage{color}
\usepackage{babel}
\usepackage{array}
\usepackage{units}
\usepackage{bm}
\usepackage{amsmath}
\usepackage{amssymb}
\usepackage{graphicx}
\usepackage[unicode=true,pdfusetitle,
 bookmarks=true,bookmarksnumbered=true,bookmarksopen=true,bookmarksopenlevel=1,
 breaklinks=false,pdfborder={0 0 0},backref=false,colorlinks=true]
 {hyperref}
\hypersetup{
 citecolor=blue,urlcolor=blue}

\makeatletter

\providecommand{\tabularnewline}{\\}

\usepackage{bm}

\usepackage[justification=centerlast,labelsep=period,font=small,figurename=Fig.]{caption}

\makeatother

\begin{document}

\title{Real and virtual $N\bar{N}$ pair production near the threshold}

\author{V. F. Dmitriev}

\email{V.F.Dmitriev@inp.nsk.su}

\author{A. I. Milstein}

\email{A.I.Milstein@inp.nsk.su}

\author{S. G. Salnikov}

\email{S.G.Salnikov@inp.nsk.su}

\affiliation{Budker Institute of Nuclear Physics, 630090 Novosibirsk, Russia}

\affiliation{Novosibirsk State University, 630090 Novosibirsk, Russia}

\date{\today}
\begin{abstract}
Nucleon-antinucleon optical potential, which explains the experimental
data for the processes $e^{+}e^{-}\rightarrow p\bar{p}$ and $e^{+}e^{-}\rightarrow\mbox{pions}$
near the threshold of $p\bar{p}$ pair production, is suggested. To
obtain this potential we have used the available experimental data
for $p\bar{p}$ scattering, $p\bar{p}$ pair production in $e^{+}e^{-}$
annihilation, and the ratio of electromagnetic form factors of a proton
in the timelike region. It turns out that final-state interaction
via the optical potential allows one to reproduce the available experimental
data with good accuracy. Our results for the cross sections of $e^{+}e^{-}\to6\pi$
process near the threshold of $p\bar{p}$ pair production are in agreement
with the recent experiments.
\end{abstract}
\maketitle
\noindent \global\long\def\sp#1{\qopname\relax{no}{Sp}#1}

\noindent \global\long\def\re#1{\qopname\relax{no}{Re}#1}

\noindent \global\long\def\im#1{\qopname\relax{no}{Im}#1}

\section{Introduction}

At present the study of nucleon-antinucleon interaction in the low-energy
region is an actual topic. Several optical nucleon-antinucleon potentials~\cite{el-bennich2009paris,zhou2012energy,Kang2014}
are usually used to describe the interaction in this region. All these
nucleon-antinucleon potentials have been proposed to fit the nucleon-antinucleon
scattering data. These data include elastic, charge-exchange, and
annihilation cross sections of $p\bar{p}$ scattering, as well as
some single-spin observables. These observables can be described very
well by any of the models~\cite{el-bennich2009paris,zhou2012energy,Kang2014}.
To discriminate different models one can use other observables. For
example, calculations were made for double-spin observables in $p\bar{p}$
scattering~\cite{dmitriev2008spin,uzikov2009forward,dmitriev2010spin,zhou2013polarization},
and the predictions of different optical potentials were indeed different.
Unfortunately, the experimental data for these observables are still
absent.

There is another set of data that one can hope to describe with the
help of potential models, namely, the cross sections of nucleon-antinucleon
production in $e^{+}e^{-}$ annihilation. It was shown in our previous
papers~\cite{dmitriev2007final,dmitriev2014isoscalar} that the cross
sections of these processes in the energy region close to the threshold
can be written in terms of the radial wave functions of nucleon-antinucleon
pair at origin. These cross sections were measured at BABAR~\cite{Lees2013},
CMD\nobreakdash-3~\cite{Akhmetshin2015} and SND~\cite{Achasov2014}.
The ratio of electromagnetic form factors of a proton in the timelike
region, that was also measured~\cite{Lees2013,Akhmetshin2015}, can
also be expressed via the wave functions. This ratio has quite strong
energy dependence near the threshold, and one needs a nontrivial model
to describe this. It was shown that some of these observables can
be described by slightly modified Paris optical potential~\cite{dmitriev2007final}
or by the Julich model~\cite{Haidenbauer2014}.

In this paper we go further and try to describe, with the help of
an optical theorem, the contribution of virtual nucleon-antinucleon
pair to the cross sections of meson production in the energy region
close to the $p\bar{p}$ threshold. We refer the cross section of
this process as inelastic cross section of nucleon-antinucleon pair
production, while the cross section of real $N\bar{N}$ pair production
is called the elastic cross section. The inelastic cross section can
be expressed in terms of the Green's function of the Schr�dinger equation
in the presence of an optical potential. First of all, we are interested
in the processes \mbox{$e^{+}e^{-}\to3\left(\pi^{+}\pi^{-}\right)$}
and \mbox{$e^{+}e^{-}\to2\left(\pi^{+}\pi^{-}\pi^{0}\right)$} because
the cross sections of $6\pi$ production have a sharp dip near the
$p\bar{p}$ threshold~\cite{Aubert2006a,Akhmetshin2013,Lukin2015},
and this phenomenon is not well understood yet. This feature is expected
to be a consequence of the interaction of virtual nucleons, because
other contributions should be smooth functions in the energy region
under consideration. There was an attempt to explain the behavior
of meson production cross sections with the help of the Julich model~\cite{Haidenbauer2015}.
Though the cross sections of some channels are reproduced, there are
a few cross sections that are not described within this model, for
instance, the cross section of the process \mbox{$e^{+}e^{-}\to3\left(\pi^{+}\pi^{-}\right)$}.
The Paris optical potential completely fails to describe the process
\mbox{$e^{+}e^{-}\to mesons$} via annihilation of virtual nucleon-antinucleon
pair. This problem appears due to very large imaginary part of the
central potential in this model. Such huge imaginary part results
in the significant overestimation of the inelastic cross section compared
with the elastic one. A strong potential appears as a result of an
attempt to describe the data in wide energy and angle regions, i.e.,
in large region of momentum transfers. The Nijmegen optical potential
has another shortcoming. This model implies that a complicated matching
condition should be applied to the wave functions at radius about
$\unit[1]{fm}$, and it is not evident how to apply this condition
to calculate the Green's function at origin.

The interaction of virtual nucleons in the process of $e^{+}e^{-}$
annihilation into mesons is very sensitive to the potential at small
distances. Unfortunately, the short-range potential can't be determined
very well from $p\bar{p}$ scattering data alone. Therefore, one should
also take into account other experimental data. This is what we perform
in the present work. We consider the partial waves with the total
angular momentum $J=1$, that contribute to the processes of nucleon-antinucleon
production in $e^{+}e^{-}$ annihilation, and assume that other partial
waves can be described by any of the models mentioned above. For $^{3}S_{1}$
and $^{3}D_{1}$ partial waves, coupled by the tensor forces, we propose
a new potential model based on the fit of experimental data for the
cross sections of $p\bar{p}$ scattering, as well as the cross sections
of nucleon-antinucleon pair production and the ratio of electromagnetic
form factors of the proton. The account for the tensor potential is
very important, because it is crucial for the description of the electromagnetic
form factors ratio. Our model qualitatively reproduces the features
of $6\pi$ production in $e^{+}e^{-}$ annihilation in the vicinity
of the threshold of $e^{+}e^{-}\to p\bar{p}$ process.

\raggedbottom

\section{Amplitude of the process}

It is shown in our recent paper~\cite{dmitriev2014isoscalar} that
in the non-relativistic approximation the amplitude $T_{\lambda\mu}^{I}$
of $N\bar{N}$ pair production in $e^{+}e^{-}$ annihilation near
the threshold can be presented for the certain isospin channel $I=0,\,1$
as follows (in units $4\pi\alpha/Q^{2}$, $\alpha$~is the fine structure
constant, $\hbar=c=1$): 
\begin{equation}
T_{\lambda\mu}^{I}=G_{s}^{I}\biggl\{\sqrt{2}u_{1R}^{I}(0)(\bm{e}_{\mu}\cdot\bm{\epsilon}_{\lambda}^{*})+u_{2R}^{I}(0)\left[(\bm{e}_{\mu}\cdot\bm{\epsilon}_{\lambda}^{*})-3(\hat{\bm{k}}\cdot\bm{e}_{\mu})(\hat{\bm{k}}\cdot\bm{\epsilon}_{\lambda}^{*})\right]\biggr\},\label{dress2}
\end{equation}
where $G_{s}^{I}$~is an energy-independent constant, ${\bf e}_{\mu}$~is
a virtual photon polarization vector, corresponding to the projection
of spin $J_{z}=\mu=\pm1$, $\bm{\epsilon}_{\lambda}$~is the spin-1
function of $N\bar{N}$ pair, $\lambda=\pm1,\,0$~is the projection
of spin on the nucleon momentum $\bm{k}$, and $\hat{\bm{k}}=\bm{k}/k$.
The radial wave functions $u_{nR}^{I}(r)$ and $w_{nR}^{I}(r)$, $n=1,\,2$,
are the regular solutions of the equations 
\begin{align}
 & \frac{p_{r}^{2}}{M}\chi_{n}+{\cal V}\chi_{n}=2E\chi_{n}\,,\nonumber \\
 & {\cal V}=\begin{pmatrix}V_{S}^{I} & -2\sqrt{2}\,V_{T}^{I}\\
-2\sqrt{2}\,V_{T}^{I} & \quad V_{D}^{I}-2V_{T}^{I}+{\displaystyle \frac{6}{Mr^{2}}}
\end{pmatrix},\qquad\chi_{n}=\begin{pmatrix}u_{n}^{I}\\
w_{n}^{I}
\end{pmatrix}.\label{equation}
\end{align}
Here $M$ is the proton mass, $E=k^{2}/(2M)$, $V_{S}^{I}(r)$, $V_{D}^{I}(r)$,
and $V_{T}^{I}(r)$ are the functions in the Hamiltonian $H^{I}$
of $N\bar{N}$ interaction for the isospin~$I$, 
\begin{align}
 & H^{I}=\frac{p_{r}^{2}}{M}+V_{S}^{I}(r)\delta_{L0}+V_{D}^{I}(r)\delta_{L2}+V_{T}^{I}(r)\,S_{12}\,,\nonumber \\
 & S_{12}=6\left(\bm{S}\cdot\bm{n}\right)^{2}-4\,,
\end{align}
where $\bm{S}$ is the spin operator for the spin-one system of the
produced pair, $(-p_{r}^{2})$ is the radial part of the Laplace operator,
$L$ denotes the orbital angular momentum, and $\bm{n}=\bm{r}/r$.
The asymptotic forms of the regular solutions (they have no singularities
at $r=0$) at large distances are~\cite{dmitriev2014isoscalar} 
\begin{align}
 & u_{1R}^{I}(r)=\frac{1}{2ikr}\Big[S_{11}^{I}\,e^{ikr}-e^{-ikr}\Big],\nonumber \\
 & w_{1R}^{I}(r)=-\frac{1}{2ikr}S_{12}^{I}\,e^{ikr}\,,\nonumber \\
 & u_{2R}^{I}(r)=\frac{1}{2ikr}S_{21}^{I}\,e^{ikr}\,,\nonumber \\
 & w_{2R}^{I}(r)=\frac{1}{2ikr}\Big[-S_{22}^{I}e^{ikr}+e^{-ikr}\Big],\label{asvuw1}
\end{align}
where $S_{ij}^{I}$ are some functions of energy, $S_{21}^{I}=S_{12}^{I}$,
$|S_{11}^{I}|^{2}+|S_{12}^{I}|^{2}\leq1$, and $|S_{22}^{I}|^{2}+|S_{21}^{I}|^{2}\leq1$.
For our purpose we also need to know the non-regular solutions of
Eqs.~\eqref{equation} which have the asymptotic forms at large distances
\begin{align}
 & u_{1N}^{I}(r)=\frac{1}{kr}\,e^{ikr}\,, &  & \lim_{r\to\infty}rw_{1N}^{I}(r)=0\,,\nonumber \\
 & \lim_{r\to\infty}ru_{2N}^{I}(r)=0\,, &  & w_{2N}^{I}(r)=-\frac{1}{kr}\,e^{ikr}\,.\label{asvuw2}
\end{align}

\section{Cross section and the Sachs form factors}

Performing summation over the polarization of nucleon pair and averaging
over the polarization of virtual photon by means of the equations,
\begin{equation}
\sum_{\lambda=1,2,3}\epsilon_{\lambda}^{i*}\epsilon_{\lambda}^{j}=\delta^{ij}\,,\qquad\frac{1}{2}\sum_{\mu=1,2}e_{\mu}^{i*}e_{\mu}^{j}=\frac{1}{2}\delta_{\perp}^{ij}=\frac{1}{2}\left(\delta^{ij}-P^{i}P^{j}/P^{2}\right),
\end{equation}
where $\bm{P}$ is the electron momentum, we obtain the cross section
corresponding to the amplitude~\eqref{dress2} in the center-of-mass
frame (see, e.g., Ref.~\cite{landau4}) 
\begin{equation}
\frac{d\sigma^{I}}{d\Omega}=\frac{\beta\alpha^{2}}{4Q^{2}}\,\left[\left|G_{M}^{I}(Q^{2})\right|^{2}(1+\cos^{2}\theta)+\frac{4M^{2}}{Q^{2}}\,\left|G_{E}^{I}(Q^{2})\right|^{2}\sin^{2}\theta\right].
\end{equation}
Here $\beta=k/M$, $Q=2(M+E)$, and $\theta$~is the angle between
the electron (positron) momentum $\bm{P}$ and the momentum of the
final particle $\bm{k}$. In terms of the form factor~$G_{s}^{I}$,
the electromagnetic Sachs form factors have the form 
\begin{align}
 & G_{M}^{I}=G_{s}^{I}\left[u_{1R}^{I}(0)+\frac{1}{\sqrt{2}}u_{2R}^{I}(0)\right],\nonumber \\
 & \frac{2M}{Q}G_{E}^{I}=G_{s}^{I}\left[u_{1R}^{I}(0)-\vphantom{\frac{1}{\sqrt{2}}}\sqrt{2}u_{2R}^{I}(0)\right].\label{eq:GeGm}
\end{align}
Thus, in the non-relativistic approximation the ratio $G_{E}^{I}/G_{M}^{I}$
is independent on the constant~$G_{s}^{I}$, 
\begin{equation}
\frac{G_{E}^{I}}{G_{M}^{I}}=\dfrac{u_{1R}^{I}(0)-\sqrt{2}u_{2R}^{I}(0)}{u_{1R}^{I}(0)+\dfrac{1}{\sqrt{2}}u_{2R}^{I}(0)}\,.
\end{equation}
Note that the electromagnetic interaction is important only in the
narrow region $\beta\sim\pi\alpha$ where the nucleon energy is $E=M\beta^{2}/2\sim\unit[0.3]{MeV}$.
In this paper we do not consider this narrow region and neglect the
electromagnetic interaction in the potential. The contribution of
the isospin~$I$ to the total cross section of the nucleon pair production
(the elastic cross section) reads 
\begin{equation}
\sigma^{I}=\frac{2\pi\beta\alpha^{2}}{Q^{2}}\left|G_{s}^{I}\right|^{2}\left[\left|u_{1R}^{I}(0)\right|^{2}+\left|u_{2R}^{I}(0)\right|^{2}\right].
\end{equation}
Thus, to describe the energy dependence of the ratio $G_{E}^{I}/G_{M}^{I}$
and the cross section $\sigma^{I}$ in the non-relativistic approximation,
it is necessary to know the functions $u_{1}^{I}(0)$ and~$u_{2}^{I}(0)$.

In~order to describe the total cross section, a sum of elastic and
inelastic cross sections (the production of mesons via annihilation
of virtual $N\bar{N}$ pair), we use the method of the Green's function.
Let us introduce the Green's function ${\cal D}(r,\,r'|E)$, 
\begin{equation}
\left(\frac{p_{r}^{2}}{M}+{\cal V}-2E\right){\cal D}\left(r,\,r'|E\right)=-\frac{1}{rr'}\delta\left(r-r'\right).\label{eq:GreenEq}
\end{equation}
Then the total cross section, $\sigma_{\mathrm{tot}}^{I}$ can be
written as~\cite{Fadin1987} 
\begin{equation}
\sigma_{\mathrm{tot}}^{I}=-\frac{2\pi\alpha^{2}}{M^{2}Q^{2}}\,\left|G_{s}^{I}\right|^{2}\,\sp{\left[\vphantom{\Bigl(\Bigr)}\im{{\cal D}\left(0,\,0|E\right)}\right]}.\label{CStot}
\end{equation}

The solution of Eq.~\eqref{eq:GreenEq} can be written in the form
\begin{equation}
{\cal D}\left(r,\,r'|E\right)=-Mk\sum_{n=1,2}\left[\vphantom{\Bigl(\Bigr)}\vartheta\left(r'-r\right)\chi_{nR}(r)\chi_{nN}^{T}(r')+\vartheta\left(r-r'\right)\chi_{nN}(r)\chi_{nR}^{T}(r')\right],\label{GF1}
\end{equation}
where $\chi^{T}$ denotes transposition of $\chi$, if the following
relations hold:
\begin{align}
 & \sum_{n=1,2}\left[\vphantom{\Bigl(\Bigr)}\chi_{nR}(r)\chi_{nN}^{T}(r)-\chi_{nN}(r)\chi_{nR}^{T}(r)\right]=\mathbf{0}\,,\nonumber \\
 & \sum_{n=1,2}\left[\vphantom{\Bigl(\Bigr)}\chi_{nR}'(r)\chi_{nN}^{T}(r)-\chi_{nN}'(r)\chi_{nR}^{T}(r)\right]=\frac{1}{kr^{2}}\mathbf{1}\,.\label{eq:GreenEq2}
\end{align}
Here $\chi'(r)=\partial\chi(r)/\partial r$, $\mathbf{0}$ and $\mathbf{1}$
stand for zero and unit matrix, respectively. The validity of Eq.~\eqref{eq:GreenEq2}
is a consequence of the relations
\begin{align}
 & \chi_{1R}^{T}(r)\chi_{2N}'(r)-\chi_{2N}^{T}(r)\chi_{1R}'(r)=0\,, &  & \chi_{2R}^{T}(r)\chi_{1N}'(r)-\chi_{1N}^{T}(r)\chi_{2R}'(r)=0\,,\nonumber \\
 & \chi_{1N}^{T}(r)\chi_{1R}'(r)-\chi_{1R}^{T}(r)\chi_{1N}'(r)=\frac{1}{kr^{2}}\,, &  & \chi_{2N}^{T}(r)\chi_{2R}'(r)-\chi_{2R}^{T}(r)\chi_{2N}'(r)=\frac{1}{kr^{2}}\,,
\end{align}
following from Eq.~\eqref{equation}, symmetry of the matrix ${\cal V}$
in that equation, and the asymptotic forms~\eqref{asvuw1} and~\eqref{asvuw2}.

\section{Results of the calculations}

We propose a simple potential model to describe the nucleon-antinucleon
interaction in the state with the total angular momentum $J=1$. This
is the only state that contributes to the processes of nucleon-antinucleon
production in $e^{+}e^{-}$ annihilation. The interaction in other
partial waves can be described very well by the models~\cite{el-bennich2009paris,zhou2012energy,Kang2014}.
The optical potential of nucleon-antinucleon interaction in Eq.~\eqref{eq:potdecomp}
can be written as
\begin{equation}
V_{n}(r)=V_{n0}(r)+V_{n1}(r)\left(\bm{\tau}_{1}\cdot\bm{\tau}_{2}\right)\,,\qquad n=S,\,D,\,T\,,\label{eq:potdecomp}
\end{equation}
where $\bm{\tau}_{i}$~are the Pauli matrices in the isospin space,
so that the potentials corresponding to $I=0,\,1$ channels read
\begin{equation}
V_{n}^{0}(r)=V_{n0}(r)-3V_{n1}(r)\,,\qquad V_{n}^{1}(r)=V_{n0}(r)+V_{n1}(r)\,.
\end{equation}
We use a potential which is the sum of a long-range pion-exchange
potential and a short-range potential well
\begin{align}
 & V_{n0}(r)=\left(U_{n0}-i\,W_{n0}\right)\theta\left(a_{n0}-r\right),\nonumber \\
 & V_{n1}(r)=\left(U_{n1}-i\,W_{n1}\right)\theta\left(a_{n1}-r\right)+\tilde{V}_{n}(r)\theta\left(r-a_{n1}\right),\label{eq:well}
\end{align}
where $\theta(x)$~is the Heaviside function, $\tilde{V}_{n}(r)$~is
the pion-exchange potential, $U_{nI}$, $W_{nI}$, and~$a_{nI}$
are free parameters fixed by fitting the experimental data. The pion-exchange
potential of nucleon-antinucleon interaction for the total spin $S=1$
is given by the expression (see,~f.i.,~\cite{Ericson1988})
\begin{align}
 & \tilde{V}_{S}(r)=\tilde{V}_{D}(r)=-f_{\pi}^{2}\frac{e^{-m_{\pi}r}}{3r}\,,\nonumber \\
 & \tilde{V}_{T}(r)=-f_{\pi}^{2}\left(\frac{1}{3}+\frac{1}{m_{\pi}r}+\frac{1}{\left(m_{\pi}r\right)^{2}}\right)\frac{e^{-m_{\pi}r}}{r}\,,
\end{align}
where $f_{\pi}^{2}=0.075$, $m_{\pi}$ is the pion mass. At small~$r$
the tensor potentials $V_{T}^{I}$ are regularized by the factor
\[
F(r)=\frac{\left(cr\right)^{2}}{1+\left(cr\right)^{2}}
\]
with $c=\unit[10]{fm^{-1}}$. Our analysis shows that one can take
the radii of real and imaginary parts of the potentials~\eqref{eq:well}
to be the same.

The electromagnetic form factors of the proton and neutron are expressed
via the isoscalar and isovector form factors~\eqref{eq:GeGm} by
the relations
\begin{align}
 & G_{E}^{p}=\frac{G_{E}^{0}+G_{E}^{1}}{\sqrt{2}}\,, &  & G_{E}^{n}=\frac{G_{E}^{0}-G_{E}^{1}}{\sqrt{2}}\,,\nonumber \\
 & G_{M}^{p}=\frac{G_{M}^{0}+G_{M}^{1}}{\sqrt{2}}\,, &  & G_{M}^{n}=\frac{G_{M}^{0}-G_{M}^{1}}{\sqrt{2}}\,.
\end{align}
Thus, the cross sections of nucleon-antinucleon production read
\begin{align}
 & \sigma^{p\bar{p}}=\frac{\pi\beta\alpha^{2}}{Q^{2}}\left[\left|G_{S}^{0}u_{1R}^{0}(0)+G_{S}^{1}u_{1R}^{1}(0)\right|^{2}+\left|G_{S}^{0}u_{2R}^{0}(0)+G_{S}^{1}u_{2R}^{1}(0)\right|^{2}\right],\nonumber \\
 & \sigma^{n\bar{n}}=\frac{\pi\beta\alpha^{2}}{Q^{2}}\left[\left|G_{S}^{0}u_{1R}^{0}(0)-G_{S}^{1}u_{1R}^{1}(0)\right|^{2}+\left|G_{S}^{0}u_{2R}^{0}(0)-G_{S}^{1}u_{2R}^{1}(0)\right|^{2}\right],
\end{align}
and the ratio of electromagnetic form factors of the proton is given~by
\begin{equation}
\frac{G_{E}^{p}}{G_{M}^{p}}=\dfrac{G_{S}^{0}u_{1R}^{0}(0)+G_{S}^{1}u_{1R}^{1}(0)-\sqrt{2}\left[\vphantom{\bigl|\bigr|}G_{S}^{0}u_{2R}^{0}(0)+G_{S}^{1}u_{2R}^{1}(0)\right]}{G_{S}^{0}u_{1R}^{0}(0)+G_{S}^{1}u_{1R}^{1}(0)+\dfrac{1}{\sqrt{2}}\left[\vphantom{\bigl|\bigr|}G_{S}^{0}u_{2R}^{0}(0)+G_{S}^{1}u_{2R}^{1}(0)\right]}\,.
\end{equation}

\begin{figure}
\begin{centering}
\includegraphics[height=5.45cm]{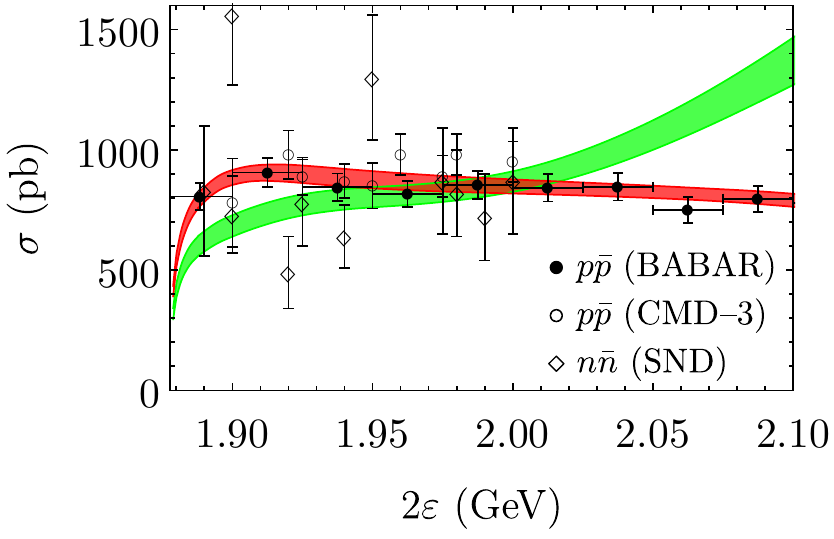}\hfill{}\includegraphics[height=5.45cm]{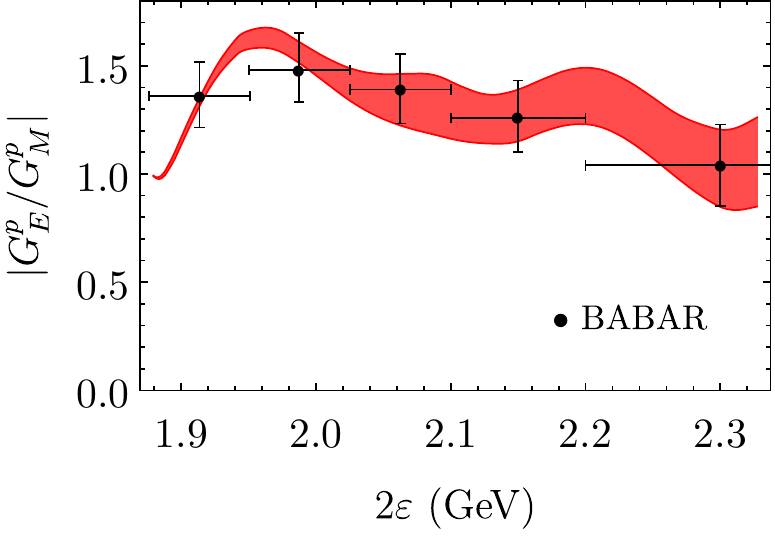}
\par\end{centering}

\caption{\label{fig:production}The cross sections of $p\bar{p}$ (red/dark
line) and $n\bar{n}$ (green/light line) production (left) and the
ratio of electromagnetic form factors of the proton (right) as a function
of total energy $2\varepsilon=2M+2E$. The experimental data are from
Refs.~\cite{Lees2013,Akhmetshin2015,Achasov2014}.\hspace*{\fill}}
\end{figure}

\begin{figure}
\begin{centering}
\includegraphics[height=5.45cm]{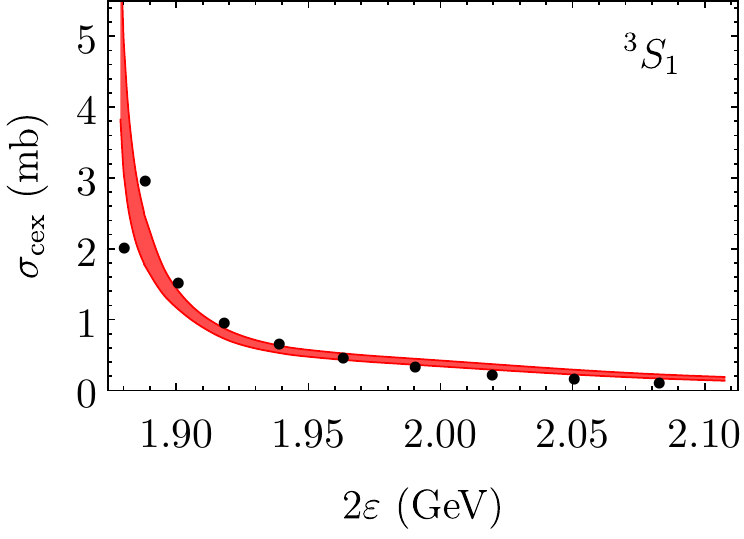}\hfill{}\includegraphics[height=5.45cm]{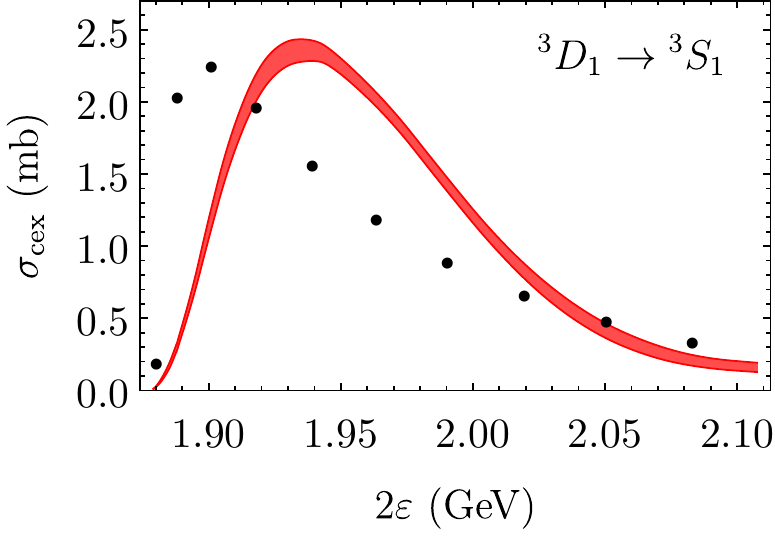}
\par\end{centering}

\caption{\label{fig:cex}$^{3}S_{1}$ and $^{3}D_{1}\to{}^{3}S_{1}$ contributions
to the charge-exchange cross section compared with the Nijmegen data~\cite{zhou2012energy},
$\varepsilon=M+E$.\hspace*{\fill}}
\end{figure}

The data used for fitting the parameters of the potential include
the cross sections of $p\bar{p}$ and $n\bar{n}$ production~\cite{Lees2013,Akhmetshin2015,Achasov2014},
the ratio of electromagnetic form factors of the proton~\cite{Lees2013}
and the partial contributions of $J=1$ waves to the elastic, change-exchange
and total cross sections of $p\bar{p}$ scattering. The partial cross
sections of $p\bar{p}$ scattering were calculated from the Nijmegen
partial wave $S$\nobreakdash-matrix (Tables~VI,~VII of Ref.~\cite{zhou2012energy}).
The results of the fit are shown in Table~\ref{tab:fit}. The accuracy
of the fit can be seen from Figs.~\ref{fig:production}--\ref{fig:cex}.

The number of free parameters in our model is $N_{\mathrm{fp}}=20$.
The total number of experimental data points for the cross sections
of $p\bar{p}$ and $n\bar{n}$ production and for the ratio $\left|G_{E}^{p}/G_{M}^{p}\right|$
is $N_{\mathrm{dat}}=35$. Thus, we have $N_{\mathrm{df}}=N_{\mathrm{dat}}-N_{\mathrm{fp}}=15$
degrees of freedom. The minimum $\chi^{2}$ per degree of freedom
is $\chi_{\mathrm{min}}^{2}/N_{\mathrm{df}}=29/15$, and is rather
large. However, large $\chi_{\mathrm{min}}^{2}$ value is originated
mainly from pour accuracy of some data points for $n\bar{n}$ production
cross section. Excluding two less accurate data points gives $\chi_{\mathrm{min}}^{2}/N_{\mathrm{df}}=16/13$,
which is good enough. The errors in Table~\ref{tab:fit} correspond
to the values of the parameters that give $\chi^{2}=\chi_{\mathrm{min}}^{2}+1$.

\begin{table}
\begin{centering}
\begin{tabular*}{16.4cm}{@{\extracolsep{\fill}}|>{\raggedright}m{2cm}|l|l|l|l|l|l|}
\hline 
 & $\qquad V_{S0}$ & $\quad V_{D0}$ & $\quad V_{T0}$ & $\quad V_{S1}$ & $\quad V_{D1}$ & $\quad V_{T1}$\tabularnewline
\hline 
$\;U\,(\mathrm{MeV})$ & $-433\pm3$ & $-140_{-36}^{+40}$ & $\hphantom{0.}58\pm4$ & $\hphantom{0}2.4_{-0.6}^{+0.7}$ & $\hphantom{.}798_{-140}^{+165}$ & $\hphantom{-}7.1\pm0.1$\tabularnewline
$\;W\,(\mathrm{MeV})$ & $\hphantom{-}224\pm10$ & $\hphantom{-14}0_{-27}^{+28}$ & $\hphantom{0.}19\pm1$ & $\hphantom{0.0}0_{-1.3}^{+0.9}$ & $\hphantom{.}456_{-107}^{+215}$ & $-0.3\pm0.3$\tabularnewline
$\;a\,(\mathrm{fm})$ & $0.564\pm0.002$ & $\hphantom{0}1.02_{-0.09}^{+0.06}$ & $1.03\pm0.02$ & $1.86_{-0.09}^{+0.08}$ & $0.49_{-0.02}^{+0.04}$ & $\hphantom{-}2.4\pm0.02$\tabularnewline
\hline 
$\;G_{S}$ & \multicolumn{3}{c|}{$G_{S}^{0}=0.179\pm0.006$} & \multicolumn{3}{c|}{$G_{S}^{1}=0.044+0.29\,i\pm0.014$}\tabularnewline
\hline 
\end{tabular*}
\par\end{centering}

\caption{\label{tab:fit}The results of the fit for the short-range potential~\eqref{eq:well}
and the constants~$G_{S}^{I}$.\hspace*{\fill}}
\end{table}

\begin{figure}
\begin{centering}
\includegraphics[height=5.45cm]{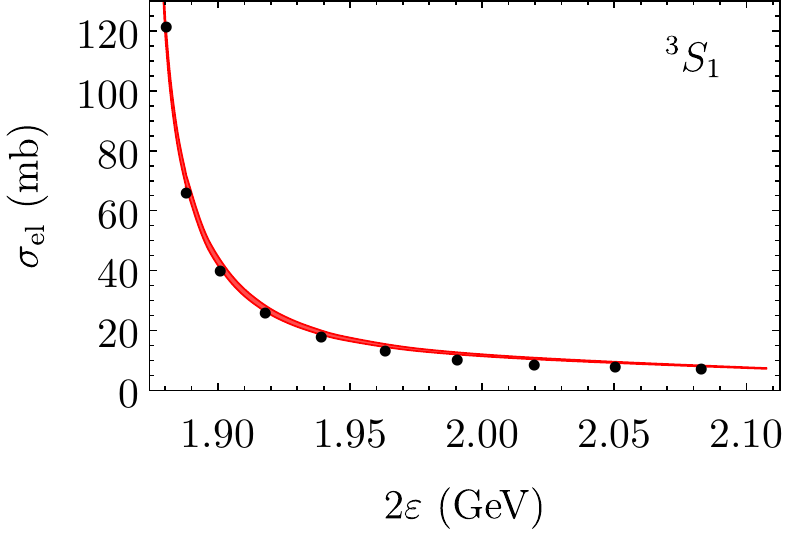}\hfill{}\includegraphics[height=5.45cm]{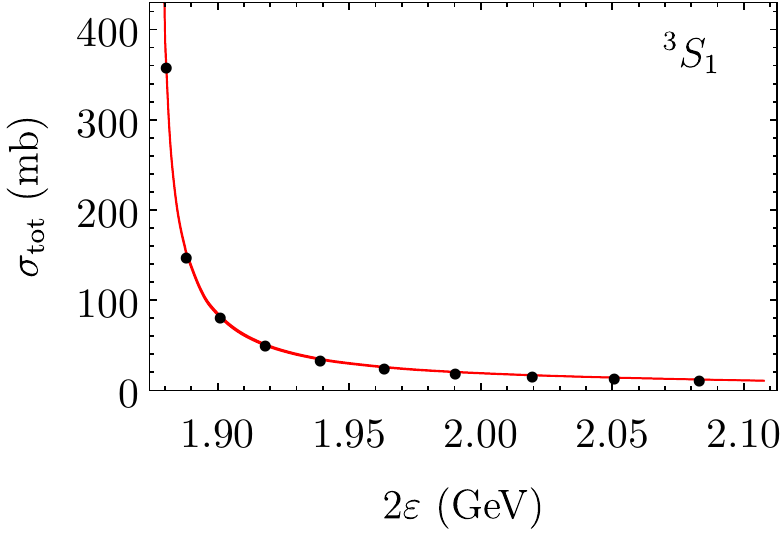}
\par\end{centering}

\begin{centering}
\includegraphics[height=5.45cm]{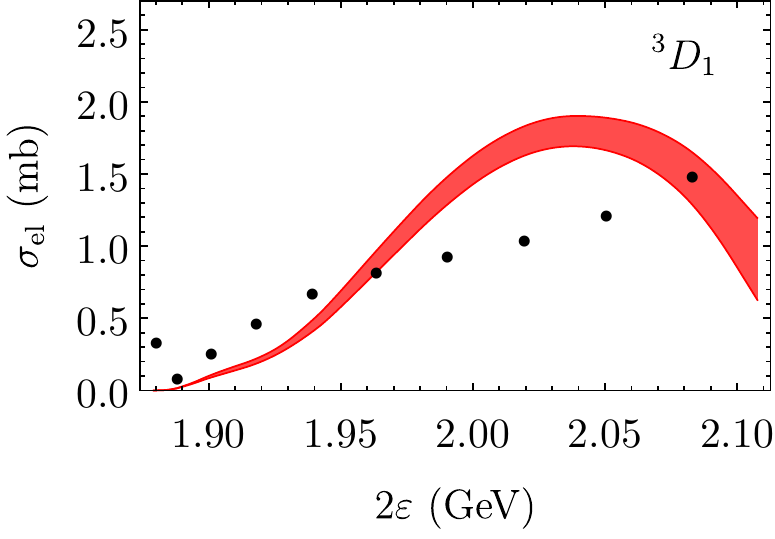}\hfill{}\includegraphics[height=5.45cm]{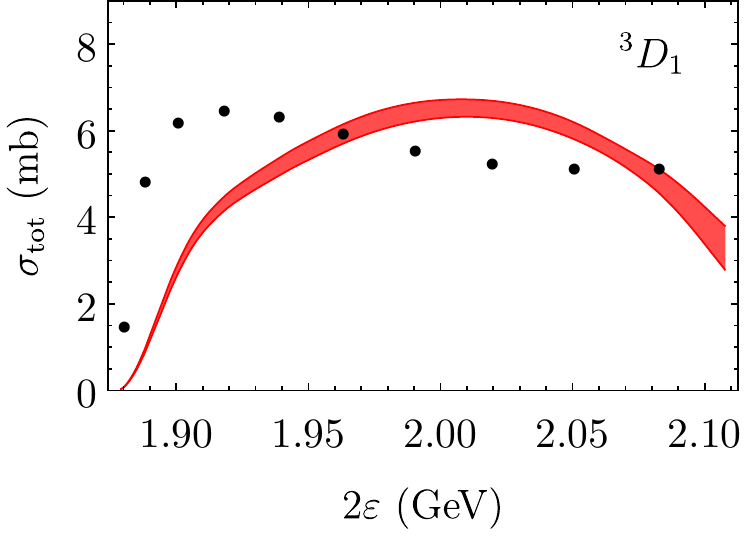}
\par\end{centering}

\caption{$^{3}S_{1}$ (first row) and $^{3}D_{1}$ (second row) contributions
to the elastic and total cross sections of $p\bar{p}$ scattering
compared with the Nijmegen data~\cite{zhou2012energy}, $\varepsilon=M+E$.\hspace*{\fill}}
\end{figure}

\begin{figure}
\begin{centering}
\includegraphics[height=5.45cm]{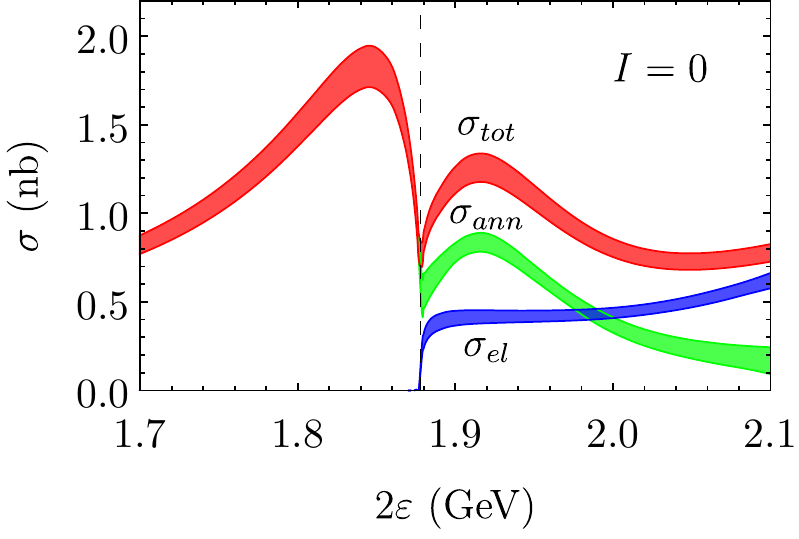}\hfill{}\includegraphics[height=5.45cm]{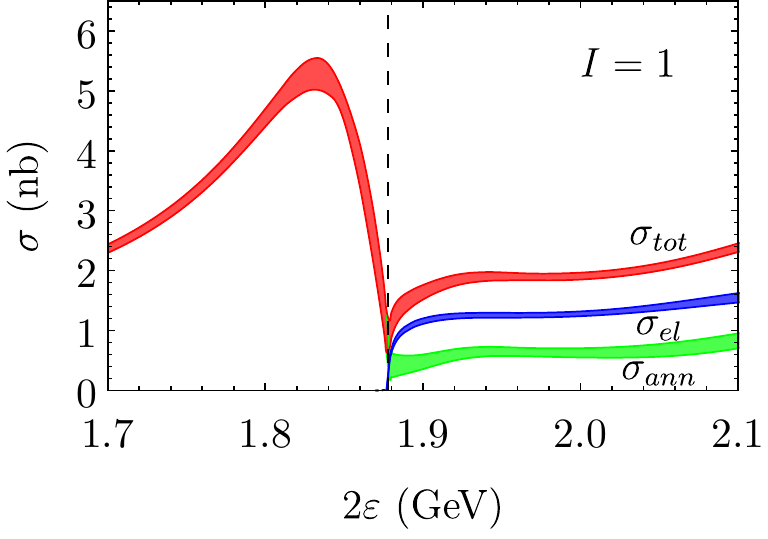}
\par\end{centering}

\caption{\label{fig:crosssections}Elastic (blue/dark line), inelastic (green/light
line) and total (red/medium line) cross sections of $e^{+}e^{-}$
annihilation through nucleon-antinucleon intermediate states with
different isospins, $\varepsilon=M+E$.\hspace*{\fill}}
\end{figure}

As soon as the potential is determined, we can calculate the Green's
function and the total cross section of pion production through nucleon-antinucleon
intermediate state~\eqref{CStot}. The elastic cross section of $N\bar{N}$
production, the total cross section and the cross section of annihilation
into mesons for different isospins are shown in Fig.~\ref{fig:crosssections}.
A dip in the total cross section of $e^{+}e^{-}$ annihilation into
mesons is predicted close to the $N\bar{N}$ threshold. This behavior
seems to be the consequence of some quasi-bound $N\bar{N}$ state
near the threshold. To check this hypothesis we have searched for
bound states in the potential considered. Our analysis shows that
there are no near-threshold bound states in the $I=1$ channel. However,
we have found a state with energy $E_{B}=\unit[\left(10-i\,32\right)]{MeV}$
in the $I=0$ channel. This state is located above the $N\bar{N}$
threshold, but it moves to $E_{B}=\unit[-21]{MeV}$ if the imaginary
part of the potential is turned off. This is an unstable bound state
in the terminology of Ref.~\cite{Badalyan1982}. This result is quite
similar to the result obtained in Ref.~\cite{Kang2014}, where $\unit[4.8]{MeV}<\re{E_{B}}<\unit[21.3]{MeV}$
and $\unit[-74.9]{MeV}<\im{E_{B}}<\unit[-60.6]{MeV}$ in the $I=0$
channel.

\begin{figure}
\begin{centering}
\includegraphics[height=5.45cm]{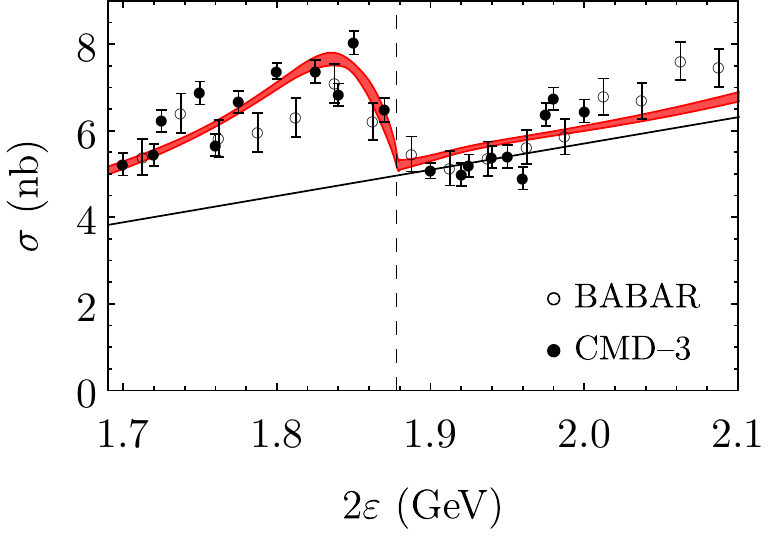}
\par\end{centering}

\caption{\label{fig:pions}The prediction for the cross section of $6\pi$
production (red thick line). The thin line shows the contribution
of non-$N\bar{N}$ channels. The data for total $6\pi$ production
are calculated from BABAR~\cite{Aubert2006a} and CMD-3~\cite{Akhmetshin2013,Lukin2015}
data. $3\left(\pi^{+}\pi^{-}\right)$ and $2\left(\pi^{+}\pi^{-}\pi^{0}\right)$
channels are taken into account, $\varepsilon=M+E$.\hspace*{\fill}}
\end{figure}

The total contribution of nucleon-antinucleon intermediate states
to the cross section of $e^{+}e^{-}$ annihilation is given by the
sum of $I=0$ and $I=1$ terms. The states with $I=0$ contribute
to the production of odd number of pions, while the states with $I=1$
contribute to the production of even number of pions. However, we
don't know accurately the pion multiplicity distribution in meson
production. This distribution was analyzed for $p\bar{p}$ annihilation
at rest~\cite{Amsler2003,Klempt2005}, and the cross section of six
pion production gives about $55\%$ of the total cross section with
$I=1$. We fit the cross section of $6\pi$ production in the energy
region between $\unit[1.7]{GeV}$ and $\unit[2.1]{GeV}$ with the
sum of inelastic $I=1$ contribution and a linear function describing
the contribution of other intermediate states (see Fig.~\ref{fig:pions}).
We obtain the best coincidence when the contribution of $6\pi$ events
to the inelastic cross section is $56\%$, which is very close to
the expectation.

\section{Conclusions}

We have proposed an optical potential describing simultaneously the
experimental data for $N\bar{N}$ scattering and $e^{+}e^{-}$ annihilation
to $N\bar{N}$ and $6\pi$ close to the threshold of $N\bar{N}$ production.
Our model predicts the dip in the total cross section of $e^{+}e^{-}$
annihilation to mesons, which is consistent with the observed behavior
of the cross section of $6\pi$ production (see Fig.~\ref{fig:pions}).
The calculation of the inelastic cross section of the process $e^{+}e^{-}\to mesons$
is based on the use of the Green's function method. The Green's function
of the Schr�dinger equation in the optical potential is derived with
the tensor forces taken into account.

It is worth noting that we found several sets of potential parameters
that fit the experimental data for $N\bar{N}$ scattering and for
the cross  sections $e^{+}e^{-}\to N\bar{N}$ with good $\chi^{2}$.
However, the account for the data for the cross section of $e^{+}e^{-}\to6\pi$
close to the threshold of $N\bar{N}$ production leads to the unique
set of parameters presented in the Table~\ref{tab:fit}. Diminishing
of uncertainties of experimental data for the cross sections of another
channels of $e^{+}e^{-}$ annihilation into mesons would be very important
for better determination of the optical potential.
\begin{acknowledgments}
This work was supported in part by RFBR under Grants No. 14\nobreakdash-02\nobreakdash-00016
and 15\nobreakdash-02\nobreakdash-07893.
\end{acknowledgments}


\begin{thebibliography}{23}%
\makeatletter
\providecommand \@ifxundefined [1]{%
 \@ifx{#1\undefined}
}%
\providecommand \@ifnum [1]{%
 \ifnum #1\expandafter \@firstoftwo
 \else \expandafter \@secondoftwo
 \fi
}%
\providecommand \@ifx [1]{%
 \ifx #1\expandafter \@firstoftwo
 \else \expandafter \@secondoftwo
 \fi
}%
\providecommand \natexlab [1]{#1}%
\providecommand \  [1]{``#1''}%
\providecommand \bibnamefont  [1]{#1}%
\providecommand \bibfnamefont [1]{#1}%
\providecommand \citenamefont [1]{#1}%
\providecommand \href@noop [0]{\@secondoftwo}%
\providecommand \href [0]{\begingroup \@sanitize@url \@href}%
\providecommand \@href[1]{\@@startlink{#1}\@@href}%
\providecommand \@@href[1]{\endgroup#1\@@endlink}%
\providecommand \@sanitize@url [0]{\catcode `\\12\catcode `\$12\catcode
  `\&12\catcode `\#12\catcode `\^12\catcode `\_12\catcode `\%12\relax}%
\providecommand \@@startlink[1]{}%
\providecommand \@@endlink[0]{}%
\providecommand \url  [0]{\begingroup\@sanitize@url \@url }%
\providecommand \@url [1]{\endgroup\@href {#1}{\urlprefix }}%
\providecommand \urlprefix  [0]{URL }%
\providecommand \Eprint [0]{\href }%
\providecommand \doibase [0]{http://dx.doi.org/}%
\providecommand \selectlanguage [0]{\@gobble}%
\providecommand \bibinfo  [0]{\@secondoftwo}%
\providecommand \bibfield  [0]{\@secondoftwo}%
\providecommand \translation [1]{[#1]}%
\providecommand \BibitemOpen [0]{}%
\providecommand \bibitemStop [0]{}%
\providecommand \bibitemNoStop [0]{.\EOS\space}%
\providecommand \EOS [0]{\spacefactor3000\relax}%
\providecommand \BibitemShut  [1]{\csname bibitem#1\endcsname}%
\let\auto@bib@innerbib\@empty
\bibitem [{\citenamefont {El-Bennich}\ \emph {et~al.}(2009)\citenamefont
  {El-Bennich}, \citenamefont {Lacombe}, \citenamefont {Loiseau},\ and\
  \citenamefont {Wycech}}]{el-bennich2009paris}%
  \BibitemOpen
\bibfield  {author} {\bibinfo {author} {\bibfnamefont {B.}~\bibnamefont
  {El-Bennich}}, \bibnamefont {et~al.},\ }\href {\doibase
  10.1103/PhysRevC.79.054001} {\bibfield  {journal} {\bibinfo  {journal} {Phys.
  Rev. C}\ }\textbf {\bibinfo {volume} {79}},\ \bibinfo {pages} {54001}
  (\bibinfo {year} {2009})}\BibitemShut {NoStop}%
\bibitem [{\citenamefont {Zhou}\ and\ \citenamefont
  {Timmermans}(2012)}]{zhou2012energy}%
  \BibitemOpen
\bibfield  {author} {\bibinfo {author} {\bibfnamefont {D.}~\bibnamefont
  {Zhou}}\ \bibnamefont {and}\ \bibinfo {author} {\bibfnamefont {R.~G.~E.}\
  \bibnamefont {Timmermans}},\ }\href {\doibase 10.1103/PhysRevC.86.044003}
  {\bibfield  {journal} {\bibinfo  {journal} {Phys. Rev. C}\ }\textbf {\bibinfo
  {volume} {86}},\ \bibinfo {pages} {44003} (\bibinfo {year}
  {2012})}\BibitemShut {NoStop}%
\bibitem [{\citenamefont {Kang}\ \emph {et~al.}(2014)\citenamefont {Kang},
  \citenamefont {Haidenbauer},\ and\ \citenamefont {Mei{\ss}ner}}]{Kang2014}%
  \BibitemOpen
\bibfield  {author} {\bibinfo {author} {\bibfnamefont {X.-W.}\ \bibnamefont
  {Kang}}, \bibinfo {author} {\bibfnamefont {J.}~\bibnamefont {Haidenbauer}},\
  \bibnamefont {and}\ \bibinfo {author} {\bibfnamefont {U.-G.}\ \bibnamefont
  {Mei{\ss}ner}},\ }\href {\doibase 10.1007/JHEP02(2014)113} {\bibfield
  {journal} {\bibinfo  {journal} {J. High Energy Phys.}\ }\textbf {\bibinfo
  {volume} {2014}},\ \bibinfo {pages} {113} (\bibinfo {year}
  {2014})}\BibitemShut {NoStop}%
\bibitem [{\citenamefont {Dmitriev}\ \emph {et~al.}(2008)\citenamefont
  {Dmitriev}, \citenamefont {Milstein},\ and\ \citenamefont
  {Strakhovenko}}]{dmitriev2008spin}%
  \BibitemOpen
\bibfield  {author} {\bibinfo {author} {\bibfnamefont {V.~F.}\ \bibnamefont
  {Dmitriev}}, \bibinfo {author} {\bibfnamefont {A.~I.}\ \bibnamefont
  {Milstein}},\ \bibnamefont {and}\ \bibinfo {author} {\bibfnamefont {V.~M.}\
  \bibnamefont {Strakhovenko}},\ }\href {\doibase 10.1016/j.nimb.2008.02.029}
  {\bibfield  {journal} {\bibinfo  {journal} {Nucl. Instruments Methods Phys.
  Res. Sect. B Beam Interact. with Mater. Atoms}\ }\textbf {\bibinfo {volume}
  {266}},\ \bibinfo {pages} {1122} (\bibinfo {year} {2008})}\BibitemShut
  {NoStop}%
\bibitem [{\citenamefont {Uzikov}\ and\ \citenamefont
  {Haidenbauer}(2009)}]{uzikov2009forward}%
  \BibitemOpen
\bibfield  {author} {\bibinfo {author} {\bibfnamefont {Y.}~\bibnamefont
  {Uzikov}}\ \bibnamefont {and}\ \bibinfo {author} {\bibfnamefont
  {J.}~\bibnamefont {Haidenbauer}},\ }\href {\doibase
  10.1103/PhysRevC.79.024617} {\bibfield  {journal} {\bibinfo  {journal} {Phys.
  Rev. C}\ }\textbf {\bibinfo {volume} {79}},\ \bibinfo {pages} {24617}
  (\bibinfo {year} {2009})}\BibitemShut {NoStop}%
\bibitem [{\citenamefont {Dmitriev}\ \emph {et~al.}(2010)\citenamefont
  {Dmitriev}, \citenamefont {Milstein},\ and\ \citenamefont
  {Salnikov}}]{dmitriev2010spin}%
  \BibitemOpen
\bibfield  {author} {\bibinfo {author} {\bibfnamefont {V.~F.}\ \bibnamefont
  {Dmitriev}}, \bibinfo {author} {\bibfnamefont {A.~I.}\ \bibnamefont
  {Milstein}},\ \bibnamefont {and}\ \bibinfo {author} {\bibfnamefont {S.~G.}\
  \bibnamefont {Salnikov}},\ }\href {\doibase 10.1016/j.physletb.2010.05.064}
  {\bibfield  {journal} {\bibinfo  {journal} {Phys. Lett. B}\ }\textbf
  {\bibinfo {volume} {690}},\ \bibinfo {pages} {427} (\bibinfo {year}
  {2010})}\BibitemShut {NoStop}%
\bibitem [{\citenamefont {Zhou}\ and\ \citenamefont
  {Timmermans}(2013)}]{zhou2013polarization}%
  \BibitemOpen
\bibfield  {author} {\bibinfo {author} {\bibfnamefont {D.}~\bibnamefont
  {Zhou}}\ \bibnamefont {and}\ \bibinfo {author} {\bibfnamefont {R.~G.~E.}\
  \bibnamefont {Timmermans}},\ }\href {\doibase 10.1103/PhysRevC.87.054005}
  {\bibfield  {journal} {\bibinfo  {journal} {Phys. Rev. C}\ }\textbf {\bibinfo
  {volume} {87}},\ \bibinfo {pages} {54005} (\bibinfo {year}
  {2013})}\BibitemShut {NoStop}%
\bibitem [{\citenamefont {Dmitriev}\ and\ \citenamefont
  {Milstein}(2007)}]{dmitriev2007final}%
  \BibitemOpen
\bibfield  {author} {\bibinfo {author} {\bibfnamefont {V.}~\bibnamefont
  {Dmitriev}}\ \bibnamefont {and}\ \bibinfo {author} {\bibfnamefont
  {A.}~\bibnamefont {Milstein}},\ }\href {\doibase
  10.1016/j.physletb.2007.06.085} {\bibfield  {journal} {\bibinfo  {journal}
  {Phys. Lett. B}\ }\textbf {\bibinfo {volume} {658}},\ \bibinfo {pages} {13}
  (\bibinfo {year} {2007})}\BibitemShut {NoStop}%
\bibitem [{\citenamefont {Dmitriev}\ \emph {et~al.}(2014)\citenamefont
  {Dmitriev}, \citenamefont {Milstein},\ and\ \citenamefont
  {Salnikov}}]{dmitriev2014isoscalar}%
  \BibitemOpen
\bibfield  {author} {\bibinfo {author} {\bibfnamefont {V.~F.}\ \bibnamefont
  {Dmitriev}}, \bibinfo {author} {\bibfnamefont {A.~I.}\ \bibnamefont
  {Milstein}},\ \bibnamefont {and}\ \bibinfo {author} {\bibfnamefont {S.~G.}\
  \bibnamefont {Salnikov}},\ }\href {\doibase 10.1134/S1063778814080043}
  {\bibfield  {journal} {\bibinfo  {journal} {Phys. At. Nucl.}\ }\textbf
  {\bibinfo {volume} {77}},\ \bibinfo {pages} {1173} (\bibinfo {year}
  {2014})}\BibitemShut {NoStop}%
\bibitem [{\citenamefont {Lees}\ \emph {et~al.}(2013)\citenamefont {Lees},
  \citenamefont {Poireau}, \citenamefont {Tisserand}, \citenamefont {Grauges},
  \citenamefont {Palano}, \citenamefont {Eigen}, \citenamefont {Stugu},
  \citenamefont {Brown}, \citenamefont {Kerth}, \citenamefont {Kolomensky},
  \citenamefont {Lynch}, \citenamefont {Koch}, \citenamefont {Schroeder},
  \citenamefont {Asgeirsson}, \citenamefont {Hearty}, \citenamefont {Mattison},
  \citenamefont {McKenna}, \citenamefont {So}, \citenamefont {Khan},
  \citenamefont {Blinov}, \citenamefont {Buzykaev}, \citenamefont {Druzhinin},
  \citenamefont {Golubev}, \citenamefont {Kravchenko}, \citenamefont {Onuchin},
  \citenamefont {Serednyakov}, \citenamefont {Skovpen}, \citenamefont
  {Solodov}, \citenamefont {Todyshev}, \citenamefont {Yushkov}, \citenamefont
  {Kirkby}, \citenamefont {Lankford}, \citenamefont {Mandelkern}, \citenamefont
  {Dey}, \citenamefont {Gary}, \citenamefont {Long}, \citenamefont {Vitug},
  \citenamefont {Campagnari}, \citenamefont {{Franco Sevilla}}, \citenamefont
  {Hong}, \citenamefont {Kovalskyi}, \citenamefont {Richman}, \citenamefont
  {West}, \citenamefont {Eisner}, \citenamefont {Lockman}, \citenamefont
  {Martinez}, \citenamefont {Schumm}, \citenamefont {Seiden}, \citenamefont
  {Chao}, \citenamefont {Cheng}, \citenamefont {Echenard}, \citenamefont
  {Flood}, \citenamefont {Hitlin}, \citenamefont {Ongmongkolkul}, \citenamefont
  {Porter}, \citenamefont {Rakitin}, \citenamefont {Andreassen}, \citenamefont
  {Huard}, \citenamefont {Meadows}, \citenamefont {Sokoloff}, \citenamefont
  {Sun}, \citenamefont {Bloom}, \citenamefont {Ford}, \citenamefont {Gaz},
  \citenamefont {Nauenberg}, \citenamefont {Smith}, \citenamefont {Wagner},
  \citenamefont {Ayad}, \citenamefont {Toki}, \citenamefont {Spaan},
  \citenamefont {Schubert}, \citenamefont {Schwierz}, \citenamefont {Bernard},
  \citenamefont {Verderi}, \citenamefont {Clark}, \citenamefont {Playfer},
  \citenamefont {Bettoni}, \citenamefont {Bozzi}, \citenamefont {Calabrese},
  \citenamefont {Cibinetto}, \citenamefont {Fioravanti}, \citenamefont
  {Garzia}, \citenamefont {Luppi}, \citenamefont {Piemontese}, \citenamefont
  {Santoro}, \citenamefont {Baldini-Ferroli}, \citenamefont {Calcaterra},
  \citenamefont {{De Sangro}}, \citenamefont {Finocchiaro}, \citenamefont
  {Patteri}, \citenamefont {Peruzzi}, \citenamefont {Piccolo}, \citenamefont
  {Rama}, \citenamefont {Zallo}, \citenamefont {Contri}, \citenamefont {Guido},
  \citenamefont {{Lo Vetere}}, \citenamefont {Monge}, \citenamefont
  {Passaggio}, \citenamefont {Patrignani}, \citenamefont {Robutti},
  \citenamefont {Bhuyan}, \citenamefont {Prasad}, \citenamefont {Morii},
  \citenamefont {Adametz}, \citenamefont {Uwer}, \citenamefont {Lacker},
  \citenamefont {Lueck}, \citenamefont {Dauncey}, \citenamefont {Mallik},
  \citenamefont {Chen}, \citenamefont {Cochran}, \citenamefont {Meyer},
  \citenamefont {Prell}, \citenamefont {Rubin}, \citenamefont {Gritsan},
  \citenamefont {Arnaud}, \citenamefont {Davier}, \citenamefont {Derkach},
  \citenamefont {Grosdidier}, \citenamefont {{Le Diberder}}, \citenamefont
  {Lutz}, \citenamefont {Malaescu}, \citenamefont {Roudeau}, \citenamefont
  {Schune}, \citenamefont {Stocchi}, \citenamefont {Wormser}, \citenamefont
  {Lange}, \citenamefont {Wright}, \citenamefont {Coleman}, \citenamefont
  {Fry}, \citenamefont {Gabathuler}, \citenamefont {Hutchcroft}, \citenamefont
  {Payne}, \citenamefont {Touramanis}, \citenamefont {Bevan}, \citenamefont
  {{Di Lodovico}}, \citenamefont {Sacco}, \citenamefont {Sigamani},
  \citenamefont {Cowan}, \citenamefont {Brown}, \citenamefont {Davis},
  \citenamefont {Denig}, \citenamefont {Fritsch}, \citenamefont {Gradl},
  \citenamefont {Griessinger}, \citenamefont {Hafner}, \citenamefont
  {Prencipe}, \citenamefont {Barlow}, \citenamefont {Lafferty}, \citenamefont
  {Behn}, \citenamefont {Cenci}, \citenamefont {Hamilton}, \citenamefont
  {Jawahery}, \citenamefont {Roberts}, \citenamefont {Dallapiccola},
  \citenamefont {Cowan}, \citenamefont {Dujmic}, \citenamefont {Sciolla},
  \citenamefont {Cheaib}, \citenamefont {Patel}, \citenamefont {Robertson},
  \citenamefont {Biassoni}, \citenamefont {Neri}, \citenamefont {Palombo},
  \citenamefont {Cremaldi}, \citenamefont {Godang}, \citenamefont {Kroeger},
  \citenamefont {Sonnek}, \citenamefont {Summers}, \citenamefont {Nguyen},
  \citenamefont {Simard}, \citenamefont {Taras}, \citenamefont {{De Nardo}},
  \citenamefont {Monorchio}, \citenamefont {Onorato}, \citenamefont {Sciacca},
  \citenamefont {Martinelli}, \citenamefont {Raven}, \citenamefont {Jessop},
  \citenamefont {Losecco}, \citenamefont {Honscheid}, \citenamefont {Kass},
  \citenamefont {Brau}, \citenamefont {Frey}, \citenamefont {Sinev},
  \citenamefont {Strom}, \citenamefont {Torrence}, \citenamefont {Feltresi},
  \citenamefont {Gagliardi}, \citenamefont {Margoni}, \citenamefont {Morandin},
  \citenamefont {Posocco}, \citenamefont {Rotondo}, \citenamefont {Simi},
  \citenamefont {Simonetto}, \citenamefont {Stroili}, \citenamefont {Akar},
  \citenamefont {Ben-Haim}, \citenamefont {Bomben}, \citenamefont {Bonneaud},
  \citenamefont {Briand}, \citenamefont {Calderini}, \citenamefont {Chauveau},
  \citenamefont {Hamon}, \citenamefont {Leruste}, \citenamefont {Marchiori},
  \citenamefont {Ocariz}, \citenamefont {Sitt}, \citenamefont {Biasini},
  \citenamefont {Manoni}, \citenamefont {Pacetti}, \citenamefont {Rossi},
  \citenamefont {Angelini}, \citenamefont {Batignani}, \citenamefont
  {Bettarini}, \citenamefont {Carpinelli}, \citenamefont {Casarosa},
  \citenamefont {Cervelli}, \citenamefont {Forti}, \citenamefont {Giorgi},
  \citenamefont {Lusiani}, \citenamefont {Oberhof}, \citenamefont {Paoloni},
  \citenamefont {Perez}, \citenamefont {Rizzo}, \citenamefont {Walsh},
  \citenamefont {{Lopes Pegna}}, \citenamefont {Olsen}, \citenamefont {Smith},
  \citenamefont {Anulli}, \citenamefont {Faccini}, \citenamefont {Ferrarotto},
  \citenamefont {Ferroni}, \citenamefont {Gaspero}, \citenamefont {{Li Gioi}},
  \citenamefont {Mazzoni}, \citenamefont {Piredda}, \citenamefont
  {B{\"{u}}nger}, \citenamefont {Gr{\"{u}}nberg}, \citenamefont {Hartmann},
  \citenamefont {Leddig}, \citenamefont {Vo{\ss}}, \citenamefont {Waldi},
  \citenamefont {Adye}, \citenamefont {Olaiya}, \citenamefont {Wilson},
  \citenamefont {Emery}, \citenamefont {{Hamel De Monchenault}}, \citenamefont
  {Vasseur}, \citenamefont {Y{\`{e}}che}, \citenamefont {Aston}, \citenamefont
  {Bard}, \citenamefont {Benitez}, \citenamefont {Cartaro}, \citenamefont
  {Convery}, \citenamefont {Dorfan}, \citenamefont {Dubois-Felsmann},
  \citenamefont {Dunwoodie}, \citenamefont {Ebert}, \citenamefont {Field},
  \citenamefont {Fulsom}, \citenamefont {Gabareen}, \citenamefont {Graham},
  \citenamefont {Hast}, \citenamefont {Innes}, \citenamefont {Kelsey},
  \citenamefont {Kim}, \citenamefont {Kocian}, \citenamefont {Leith},
  \citenamefont {Lewis}, \citenamefont {Lindemann}, \citenamefont {Lindquist},
  \citenamefont {Luitz}, \citenamefont {Luth}, \citenamefont {Lynch},
  \citenamefont {MacFarlane}, \citenamefont {Muller}, \citenamefont {Neal},
  \citenamefont {Nelson}, \citenamefont {Perl}, \citenamefont {Pulliam},
  \citenamefont {Ratcliff}, \citenamefont {Roodman}, \citenamefont {Salnikov},
  \citenamefont {Schindler}, \citenamefont {Snyder}, \citenamefont {Su},
  \citenamefont {Sullivan}, \citenamefont {Va'Vra}, \citenamefont {Wagner},
  \citenamefont {Wang}, \citenamefont {Wisniewski}, \citenamefont {Wittgen},
  \citenamefont {Wright}, \citenamefont {Wulsin}, \citenamefont {Ziegler},
  \citenamefont {Park}, \citenamefont {Purohit}, \citenamefont {White},
  \citenamefont {Wilson}, \citenamefont {Randle-Conde}, \citenamefont {Sekula},
  \citenamefont {Bellis}, \citenamefont {Burchat}, \citenamefont {Miyashita},
  \citenamefont {Puccio}, \citenamefont {Alam}, \citenamefont {Ernst},
  \citenamefont {Gorodeisky}, \citenamefont {Guttman}, \citenamefont {Peimer},
  \citenamefont {Soffer}, \citenamefont {Spanier}, \citenamefont {Ritchie},
  \citenamefont {Ruland}, \citenamefont {Schwitters}, \citenamefont {Wray},
  \citenamefont {Izen}, \citenamefont {Lou}, \citenamefont {Bianchi},
  \citenamefont {Gamba}, \citenamefont {Zambito}, \citenamefont {Lanceri},
  \citenamefont {Vitale}, \citenamefont {Martinez-Vidal}, \citenamefont
  {Oyanguren}, \citenamefont {Villanueva-Perez}, \citenamefont {Ahmed},
  \citenamefont {Albert}, \citenamefont {Banerjee}, \citenamefont
  {Bernlochner}, \citenamefont {Choi}, \citenamefont {King}, \citenamefont
  {Kowalewski}, \citenamefont {Lewczuk}, \citenamefont {Nugent}, \citenamefont
  {Roney}, \citenamefont {Sobie}, \citenamefont {Tasneem}, \citenamefont
  {Gershon}, \citenamefont {Harrison}, \citenamefont {Latham}, \citenamefont
  {Band}, \citenamefont {Dasu}, \citenamefont {Pan}, \citenamefont {Prepost},\
  and\ \citenamefont {Wu}}]{Lees2013}%
  \BibitemOpen
\bibfield  {author} {\bibinfo {author} {\bibfnamefont {J.~P.}\ \bibnamefont
  {Lees}}, \bibnamefont {et~al.},\ }\href {\doibase 10.1103/PhysRevD.87.092005}
  {\bibfield  {journal} {\bibinfo  {journal} {Phys. Rev. D}\ }\textbf {\bibinfo
  {volume} {87}},\ \bibinfo {pages} {92005} (\bibinfo {year}
  {2013})}\BibitemShut {NoStop}%
\bibitem [{\citenamefont {Akhmetshin}\ \emph {et~al.}(2015)\citenamefont
  {Akhmetshin}, \citenamefont {Amirkhanov}, \citenamefont {Anisenkov},
  \citenamefont {Aulchenko}, \citenamefont {Banzarov}, \citenamefont
  {Bashtovoy}, \citenamefont {Berkaev}, \citenamefont {Bondar}, \citenamefont
  {Bragin}, \citenamefont {Eidelman}, \citenamefont {Epifanov}, \citenamefont
  {Epshteyn}, \citenamefont {Erofeev}, \citenamefont {Fedotovich},
  \citenamefont {Gayazov}, \citenamefont {Grebenuk}, \citenamefont {Gribanov},
  \citenamefont {Grigoriev}, \citenamefont {Gromov}, \citenamefont {Ignatov},
  \citenamefont {Ivanov}, \citenamefont {Karpov}, \citenamefont {Kasaev},
  \citenamefont {Kazanin}, \citenamefont {Khazin}, \citenamefont {Kirpotin},
  \citenamefont {Koop}, \citenamefont {Kovalenko}, \citenamefont {Kozyrev},
  \citenamefont {Kozyrev}, \citenamefont {Krokovny}, \citenamefont {Kuzmenko},
  \citenamefont {Kuzmin}, \citenamefont {Logashenko}, \citenamefont {Lukin},
  \citenamefont {Mikhailov}, \citenamefont {Okhapkin}, \citenamefont {Otboev},
  \citenamefont {Pestov}, \citenamefont {Pivovarov}, \citenamefont {Popov},
  \citenamefont {Razuvaev}, \citenamefont {Romanov}, \citenamefont {Ruban},
  \citenamefont {Ryskulov}, \citenamefont {Ryzhenenkov}, \citenamefont
  {Shebalin}, \citenamefont {Shemyakin}, \citenamefont {Shwartz}, \citenamefont
  {Shwartz}, \citenamefont {Sibidanov}, \citenamefont {Shatunov}, \citenamefont
  {Shatunov}, \citenamefont {Solodov}, \citenamefont {Titov}, \citenamefont
  {Talyshev}, \citenamefont {Vorobiov}, \citenamefont {Yudin},\ and\
  \citenamefont {Zharinov}}]{Akhmetshin2015}%
  \BibitemOpen
\bibfield  {author} {\bibinfo {author} {\bibfnamefont {R.~R.}\ \bibnamefont
  {Akhmetshin}}, \bibnamefont {et~al.},\ }\emph {\ {\bibinfo {title} {{Study of
  the process $e^+e^-\to p\bar{p}$ in the c.m. energy range from threshold to 2
  GeV with the CMD-3 detector}},}\ } (\bibinfo {year} {2015}),\ \Eprint
  {http://arxiv.org/abs/1507.08013} {arXiv:1507.08013~[hep-ex]}\BibitemShut
  {NoStop}%
\bibitem [{\citenamefont {Achasov}\ \emph {et~al.}(2014)\citenamefont
  {Achasov}, \citenamefont {Barnyakov}, \citenamefont {Beloborodov},
  \citenamefont {Berdyugin}, \citenamefont {Berkaev}, \citenamefont
  {Bogdanchikov}, \citenamefont {Botov}, \citenamefont {Dimova}, \citenamefont
  {Druzhinin}, \citenamefont {Golubev}, \citenamefont {Kardapoltsev},
  \citenamefont {Kasaev}, \citenamefont {Kharlamov}, \citenamefont {Kirpotin},
  \citenamefont {Koop}, \citenamefont {Korol}, \citenamefont {Koshuba},
  \citenamefont {Kovrizhin}, \citenamefont {Kupich}, \citenamefont {Martin},
  \citenamefont {Obrazovsky}, \citenamefont {Pakhtusova}, \citenamefont
  {Rogovsky}, \citenamefont {Senchenko}, \citenamefont {Serednyakov},
  \citenamefont {Silagadze}, \citenamefont {Shatunov}, \citenamefont {Shtol},
  \citenamefont {Shwartz}, \citenamefont {Skrinsky}, \citenamefont {Surin},
  \citenamefont {Tikhonov}, \citenamefont {Usov},\ and\ \citenamefont
  {Vasiljev}}]{Achasov2014}%
  \BibitemOpen
\bibfield  {author} {\bibinfo {author} {\bibfnamefont {M.~N.}\ \bibnamefont
  {Achasov}}, \bibnamefont {et~al.},\ }\href {\doibase
  10.1103/PhysRevD.90.112007} {\bibfield  {journal} {\bibinfo  {journal} {Phys.
  Rev. D}\ }\textbf {\bibinfo {volume} {90}},\ \bibinfo {pages} {112007}
  (\bibinfo {year} {2014})}\BibitemShut {NoStop}%
\bibitem [{\citenamefont {Haidenbauer}\ \emph {et~al.}(2014)\citenamefont
  {Haidenbauer}, \citenamefont {Kang},\ and\ \citenamefont
  {Mei{\ss}ner}}]{Haidenbauer2014}%
  \BibitemOpen
\bibfield  {author} {\bibinfo {author} {\bibfnamefont {J.}~\bibnamefont
  {Haidenbauer}}, \bibinfo {author} {\bibfnamefont {X.-W.~W.}\ \bibnamefont
  {Kang}},\ \bibnamefont {and}\ \bibinfo {author} {\bibfnamefont {U.-G.~G.}\
  \bibnamefont {Mei{\ss}ner}},\ }\href {\doibase
  10.1016/j.nuclphysa.2014.06.007} {\bibfield  {journal} {\bibinfo  {journal}
  {Nucl. Phys. A}\ }\textbf {\bibinfo {volume} {929}},\ \bibinfo {pages} {102}
  (\bibinfo {year} {2014})}\BibitemShut {NoStop}%
\bibitem [{\citenamefont {Aubert}\ \emph {et~al.}(2006)\citenamefont {Aubert},
  \citenamefont {Barate}, \citenamefont {Boutigny}, \citenamefont {Couderc},
  \citenamefont {Karyotakis}, \citenamefont {Lees}, \citenamefont {Poireau},
  \citenamefont {Tisserand}, \citenamefont {Zghiche}, \citenamefont {Grauges},
  \citenamefont {Palano}, \citenamefont {Pappagallo}, \citenamefont {Chen},
  \citenamefont {Qi}, \citenamefont {Rong}, \citenamefont {Wang}, \citenamefont
  {Zhu}, \citenamefont {Eigen}, \citenamefont {Ofte}, \citenamefont {Stugu},
  \citenamefont {Abrams}, \citenamefont {Battaglia}, \citenamefont {Best},
  \citenamefont {Brown}, \citenamefont {Button-Shafer}, \citenamefont {Cahn},
  \citenamefont {Charles}, \citenamefont {Day}, \citenamefont {Gill},
  \citenamefont {Gritsan}, \citenamefont {Groysman}, \citenamefont {Jacobsen},
  \citenamefont {Kadel}, \citenamefont {Kadyk}, \citenamefont {Kerth},
  \citenamefont {Kolomensky}, \citenamefont {Kukartsev}, \citenamefont {Lynch},
  \citenamefont {Mir}, \citenamefont {Oddone}, \citenamefont {Orimoto},
  \citenamefont {Pripstein}, \citenamefont {Roe}, \citenamefont {Ronan},
  \citenamefont {Wenzel}, \citenamefont {Barrett}, \citenamefont {Ford},
  \citenamefont {Harrison}, \citenamefont {Hart}, \citenamefont {Hawkes},
  \citenamefont {Morgan}, \citenamefont {Watson}, \citenamefont {Fritsch},
  \citenamefont {Goetzen}, \citenamefont {Held}, \citenamefont {Koch},
  \citenamefont {Lewandowski}, \citenamefont {Pelizaeus}, \citenamefont
  {Peters}, \citenamefont {Schroeder}, \citenamefont {Steinke}, \citenamefont
  {Boyd}, \citenamefont {Burke}, \citenamefont {Cottingham}, \citenamefont
  {Walker}, \citenamefont {Cuhadar-Donszelmann}, \citenamefont {Fulsom},
  \citenamefont {Hearty}, \citenamefont {Knecht}, \citenamefont {Mattison},
  \citenamefont {McKenna}, \citenamefont {Khan}, \citenamefont {Kyberd},
  \citenamefont {Saleem}, \citenamefont {Teodorescu}, \citenamefont {Blinov},
  \citenamefont {Bukin}, \citenamefont {Druzhinin}, \citenamefont {Golubev},
  \citenamefont {Kravchenko}, \citenamefont {Onuchin}, \citenamefont
  {Serednyakov}, \citenamefont {Skovpen}, \citenamefont {Solodov},
  \citenamefont {Todyshev}, \citenamefont {Bondioli}, \citenamefont {Bruinsma},
  \citenamefont {Chao}, \citenamefont {Curry}, \citenamefont {Eschrich},
  \citenamefont {Kirkby}, \citenamefont {Lankford}, \citenamefont {Lund},
  \citenamefont {Mandelkern}, \citenamefont {Mommsen}, \citenamefont {Roethel},
  \citenamefont {Stoker}, \citenamefont {Abachi}, \citenamefont {Buchanan},
  \citenamefont {Foulkes}, \citenamefont {Gary}, \citenamefont {Long},
  \citenamefont {Shen}, \citenamefont {Wang}, \citenamefont {Zhang},
  \citenamefont {{Del Re}}, \citenamefont {Hadavand}, \citenamefont {Hill},
  \citenamefont {Paar}, \citenamefont {Rahatlou}, \citenamefont {Sharma},
  \citenamefont {Berryhill}, \citenamefont {Campagnari}, \citenamefont {Cunha},
  \citenamefont {Dahmes}, \citenamefont {Hong}, \citenamefont {Richman},
  \citenamefont {Beck}, \citenamefont {Eisner}, \citenamefont {Flacco},
  \citenamefont {Heusch}, \citenamefont {Kroseberg}, \citenamefont {Lockman},
  \citenamefont {Nesom}, \citenamefont {Schalk}, \citenamefont {Schumm},
  \citenamefont {Seiden}, \citenamefont {Spradlin}, \citenamefont {Williams},
  \citenamefont {Wilson}, \citenamefont {Albert}, \citenamefont {Chen},
  \citenamefont {Dubois-Felsmann}, \citenamefont {Dvoretskii}, \citenamefont
  {Hitlin}, \citenamefont {Narsky}, \citenamefont {Piatenko}, \citenamefont
  {Porter}, \citenamefont {Ryd}, \citenamefont {Samuel}, \citenamefont
  {Andreassen}, \citenamefont {Mancinelli}, \citenamefont {Meadows},
  \citenamefont {Sokoloff}, \citenamefont {Blanc}, \citenamefont {Bloom},
  \citenamefont {Chen}, \citenamefont {Ford}, \citenamefont {Hirschauer},
  \citenamefont {Kreisel}, \citenamefont {Nauenberg}, \citenamefont {Olivas},
  \citenamefont {Ruddick}, \citenamefont {Smith}, \citenamefont {Ulmer},
  \citenamefont {Wagner}, \citenamefont {Zhang}, \citenamefont {Chen},
  \citenamefont {Eckhart}, \citenamefont {Soffer}, \citenamefont {Toki},
  \citenamefont {Wilson}, \citenamefont {Winklmeier}, \citenamefont {Zeng},
  \citenamefont {Altenburg}, \citenamefont {Feltresi}, \citenamefont {Hauke},
  \citenamefont {Jasper}, \citenamefont {Spaan}, \citenamefont {Brandt},
  \citenamefont {Dickopp}, \citenamefont {Klose}, \citenamefont {Lacker},
  \citenamefont {Nogowski}, \citenamefont {Otto}, \citenamefont {Petzold},
  \citenamefont {Schubert}, \citenamefont {Schubert}, \citenamefont {Schwierz},
  \citenamefont {Sundermann}, \citenamefont {Volk}, \citenamefont {Bernard},
  \citenamefont {Bonneaud}, \citenamefont {Grenier}, \citenamefont {Latour},
  \citenamefont {Schrenk}, \citenamefont {Thiebaux}, \citenamefont
  {Vasileiadis}, \citenamefont {Verderi}, \citenamefont {Bard}, \citenamefont
  {Clark}, \citenamefont {Gradl}, \citenamefont {Muheim}, \citenamefont
  {Playfer}, \citenamefont {Xie}, \citenamefont {Andreotti}, \citenamefont
  {Bettoni}, \citenamefont {Bozzi}, \citenamefont {Calabrese}, \citenamefont
  {Cibinetto}, \citenamefont {Luppi}, \citenamefont {Negrini}, \citenamefont
  {Piemontese}, \citenamefont {Anulli}, \citenamefont {Baldini-Ferroli},
  \citenamefont {Calcaterra}, \citenamefont {{De Sangro}}, \citenamefont
  {Finocchiaro}, \citenamefont {Pacetti}, \citenamefont {Patteri},
  \citenamefont {Peruzzi}, \citenamefont {Piccolo}, \citenamefont {Zallo},
  \citenamefont {Buzzo}, \citenamefont {Capra}, \citenamefont {Contri},
  \citenamefont {Vetere}, \citenamefont {Macri}, \citenamefont {Monge},
  \citenamefont {Passaggio}, \citenamefont {Patrignani}, \citenamefont
  {Robutti}, \citenamefont {Santroni}, \citenamefont {Tosi}, \citenamefont
  {Brandenburg}, \citenamefont {Chaisanguanthum}, \citenamefont {Morii},
  \citenamefont {Wu}, \citenamefont {Dubitzky}, \citenamefont {Marks},
  \citenamefont {Schenk}, \citenamefont {Uwer}, \citenamefont {Bhimji},
  \citenamefont {Bowerman}, \citenamefont {Dauncey}, \citenamefont {Egede},
  \citenamefont {Flack}, \citenamefont {Gaillard}, \citenamefont {Nash},
  \citenamefont {Nikolich}, \citenamefont {Vazquez}, \citenamefont {Chai},
  \citenamefont {Charles}, \citenamefont {Mader}, \citenamefont {Mallik},
  \citenamefont {Ziegler}, \citenamefont {Cochran}, \citenamefont {Crawley},
  \citenamefont {Dong}, \citenamefont {Eyges}, \citenamefont {Meyer},
  \citenamefont {Prell}, \citenamefont {Rosenberg}, \citenamefont {Rubin},
  \citenamefont {Schott}, \citenamefont {Arnaud}, \citenamefont {Davier},
  \citenamefont {Grosdidier}, \citenamefont {H{\"{o}}cker}, \citenamefont
  {Diberder}, \citenamefont {Lepeltier}, \citenamefont {Lutz}, \citenamefont
  {Oyanguren}, \citenamefont {Petersen}, \citenamefont {Pruvot}, \citenamefont
  {Rodier}, \citenamefont {Roudeau}, \citenamefont {Schune}, \citenamefont
  {Stocchi}, \citenamefont {Wang}, \citenamefont {Wormser}, \citenamefont
  {Cheng}, \citenamefont {Lange}, \citenamefont {Wright}, \citenamefont
  {Bevan}, \citenamefont {Chavez}, \citenamefont {Forster}, \citenamefont
  {Fry}, \citenamefont {Gabathuler}, \citenamefont {Gamet}, \citenamefont
  {George}, \citenamefont {Hutchcroft}, \citenamefont {Payne}, \citenamefont
  {Schofield}, \citenamefont {Touramanis}, \citenamefont {Lodovico},
  \citenamefont {Menges}, \citenamefont {Sacco}, \citenamefont {Brown},
  \citenamefont {Cowan}, \citenamefont {Flaecher}, \citenamefont {Green},
  \citenamefont {Hopkins}, \citenamefont {Jackson}, \citenamefont {McMahon},
  \citenamefont {Ricciardi}, \citenamefont {Salvatore}, \citenamefont {Brown},
  \citenamefont {Davis}, \citenamefont {Allison}, \citenamefont {Barlow},
  \citenamefont {Barlow}, \citenamefont {Chia}, \citenamefont {Edgar},
  \citenamefont {Kelly}, \citenamefont {Lafferty}, \citenamefont {Naisbit},
  \citenamefont {Williams}, \citenamefont {Yi}, \citenamefont {Chen},
  \citenamefont {Hulsbergen}, \citenamefont {Jawahery}, \citenamefont
  {Kovalskyi}, \citenamefont {Lae}, \citenamefont {Roberts}, \citenamefont
  {Simi}, \citenamefont {Blaylock}, \citenamefont {Dallapiccola}, \citenamefont
  {Hertzbach}, \citenamefont {Kofler}, \citenamefont {Li}, \citenamefont
  {Moore}, \citenamefont {Saremi}, \citenamefont {Staengle}, \citenamefont
  {Willocq}, \citenamefont {Cowan}, \citenamefont {Koeneke}, \citenamefont
  {Sciolla}, \citenamefont {Sekula}, \citenamefont {Spitznagel}, \citenamefont
  {Taylor}, \citenamefont {Yamamoto}, \citenamefont {Kim}, \citenamefont
  {Patel}, \citenamefont {Potter}, \citenamefont {Robertson}, \citenamefont
  {Lazzaro}, \citenamefont {Lombardo}, \citenamefont {Palombo}, \citenamefont
  {Bauer}, \citenamefont {Cremaldi}, \citenamefont {Eschenburg}, \citenamefont
  {Godang}, \citenamefont {Kroeger}, \citenamefont {Reidy}, \citenamefont
  {Sanders}, \citenamefont {Summers}, \citenamefont {Zhao}, \citenamefont
  {Brunet}, \citenamefont {C{\^{o}}t{\'{e}}}, \citenamefont {Taras},
  \citenamefont {Viaud}, \citenamefont {Nicholson}, \citenamefont {Cavallo},
  \citenamefont {Nardo}, \citenamefont {Fabozzi}, \citenamefont {Gatto},
  \citenamefont {Lista}, \citenamefont {Monorchio}, \citenamefont {Paolucci},
  \citenamefont {Piccolo}, \citenamefont {Sciacca}, \citenamefont {Baak},
  \citenamefont {Bulten}, \citenamefont {Raven}, \citenamefont {Snoek},
  \citenamefont {Jessop}, \citenamefont {Losecco}, \citenamefont
  {Allmendinger}, \citenamefont {Benelli}, \citenamefont {Gan}, \citenamefont
  {Honscheid}, \citenamefont {Hufnagel}, \citenamefont {Jackson}, \citenamefont
  {Kagan}, \citenamefont {Kass}, \citenamefont {Pulliam}, \citenamefont
  {Rahimi}, \citenamefont {Ter-Antonyan}, \citenamefont {Wong}, \citenamefont
  {Blount}, \citenamefont {Brau}, \citenamefont {Frey}, \citenamefont
  {Igonkina}, \citenamefont {Lu}, \citenamefont {Rahmat}, \citenamefont
  {Sinev}, \citenamefont {Strom}, \citenamefont {Strube}, \citenamefont
  {Torrence}, \citenamefont {Galeazzi}, \citenamefont {Margoni}, \citenamefont
  {Morandin}, \citenamefont {Pompili}, \citenamefont {Posocco}, \citenamefont
  {Rotondo}, \citenamefont {Simonetto}, \citenamefont {Stroili}, \citenamefont
  {Voci}, \citenamefont {Benayoun}, \citenamefont {Chauveau}, \citenamefont
  {David}, \citenamefont {Buono}, \citenamefont {{De La Vaissi{\`{e}}re}},
  \citenamefont {Hamon}, \citenamefont {Hartfiel}, \citenamefont {John},
  \citenamefont {Leruste}, \citenamefont {Malcl{\`{e}}s}, \citenamefont
  {Ocariz}, \citenamefont {Roos}, \citenamefont {Therin}, \citenamefont
  {Behera}, \citenamefont {Gladney}, \citenamefont {Panetta}, \citenamefont
  {Biasini}, \citenamefont {Covarelli}, \citenamefont {Pioppi}, \citenamefont
  {Angelini}, \citenamefont {Batignani}, \citenamefont {Bettarini},
  \citenamefont {Bucci}, \citenamefont {Calderini}, \citenamefont {Carpinelli},
  \citenamefont {Cenci}, \citenamefont {Forti}, \citenamefont {Giorgi},
  \citenamefont {Lusiani}, \citenamefont {Marchiori}, \citenamefont {Mazur},
  \citenamefont {Morganti}, \citenamefont {Neri}, \citenamefont {Paoloni},
  \citenamefont {Rama}, \citenamefont {Rizzo}, \citenamefont {Walsh},
  \citenamefont {Haire}, \citenamefont {Judd}, \citenamefont {Wagoner},
  \citenamefont {Biesiada}, \citenamefont {Danielson}, \citenamefont {Elmer},
  \citenamefont {Lau}, \citenamefont {Lu}, \citenamefont {Olsen}, \citenamefont
  {Smith}, \citenamefont {Telnov}, \citenamefont {Bellini}, \citenamefont
  {Cavoto}, \citenamefont {D'Orazio}, \citenamefont {Marco}, \citenamefont
  {Faccini}, \citenamefont {Ferrarotto}, \citenamefont {Ferroni}, \citenamefont
  {Gaspero}, \citenamefont {Gioi}, \citenamefont {Mazzoni}, \citenamefont
  {Morganti}, \citenamefont {Piredda}, \citenamefont {Polci}, \citenamefont
  {Tehrani}, \citenamefont {Voena}, \citenamefont {Schr{\"{o}}der},
  \citenamefont {Waldi}, \citenamefont {Adye}, \citenamefont {Groot},
  \citenamefont {Franek}, \citenamefont {Olaiya}, \citenamefont {Wilson},
  \citenamefont {Emery}, \citenamefont {Gaidot}, \citenamefont {Ganzhur},
  \citenamefont {{De Monchenault}}, \citenamefont {Kozanecki}, \citenamefont
  {Legendre}, \citenamefont {Mayer}, \citenamefont {Vasseur}, \citenamefont
  {Y{\`{e}}che}, \citenamefont {Zito}, \citenamefont {Park}, \citenamefont
  {Purohit}, \citenamefont {Weidemann}, \citenamefont {Wilson}, \citenamefont
  {Allen}, \citenamefont {Aston}, \citenamefont {Bartoldus}, \citenamefont
  {Berger}, \citenamefont {Boyarski}, \citenamefont {Claus}, \citenamefont
  {Coleman}, \citenamefont {Convery}, \citenamefont {Cristinziani},
  \citenamefont {Dingfelder}, \citenamefont {Dong}, \citenamefont {Dorfan},
  \citenamefont {Dujmic}, \citenamefont {Dunwoodie}, \citenamefont {Field},
  \citenamefont {Glanzman}, \citenamefont {Gowdy}, \citenamefont {Halyo},
  \citenamefont {Hast}, \citenamefont {Hryn'ova}, \citenamefont {Innes},
  \citenamefont {Kelsey}, \citenamefont {Kim}, \citenamefont {Kocian},
  \citenamefont {Leith}, \citenamefont {Libby}, \citenamefont {Luitz},
  \citenamefont {Luth}, \citenamefont {Lynch}, \citenamefont {MacFarlane},
  \citenamefont {Marsiske}, \citenamefont {Messner}, \citenamefont {Muller},
  \citenamefont {O'Grady}, \citenamefont {Ozcan}, \citenamefont {Perazzo},
  \citenamefont {Perl}, \citenamefont {Ratcliff}, \citenamefont {Roodman},
  \citenamefont {Salnikov}, \citenamefont {Schindler}, \citenamefont
  {Schwiening}, \citenamefont {Snyder}, \citenamefont {Stelzer}, \citenamefont
  {Su}, \citenamefont {Sullivan}, \citenamefont {Suzuki}, \citenamefont
  {Swain}, \citenamefont {Thompson}, \citenamefont {Va'vra}, \citenamefont
  {{Van Bakel}}, \citenamefont {Weaver}, \citenamefont {Weinstein},
  \citenamefont {Wisniewski}, \citenamefont {Wittgen}, \citenamefont {Wright},
  \citenamefont {Yarritu}, \citenamefont {Yi}, \citenamefont {Young},
  \citenamefont {Burchat}, \citenamefont {Edwards}, \citenamefont {Majewski},
  \citenamefont {Petersen}, \citenamefont {Roat}, \citenamefont {Wilden},
  \citenamefont {Ahmed}, \citenamefont {Alam}, \citenamefont {Bula},
  \citenamefont {Ernst}, \citenamefont {Jain}, \citenamefont {Pan},
  \citenamefont {Saeed}, \citenamefont {Wappler}, \citenamefont {Zain},
  \citenamefont {Bugg}, \citenamefont {Krishnamurthy}, \citenamefont {Spanier},
  \citenamefont {Eckmann}, \citenamefont {Ritchie}, \citenamefont {Satpathy},
  \citenamefont {Schwitters}, \citenamefont {Izen}, \citenamefont {Kitayama},
  \citenamefont {Lou}, \citenamefont {Ye}, \citenamefont {Bianchi},
  \citenamefont {Bona}, \citenamefont {Gallo}, \citenamefont {Gamba},
  \citenamefont {Bomben}, \citenamefont {Bosisio}, \citenamefont {Cartaro},
  \citenamefont {Cossutti}, \citenamefont {Ricca}, \citenamefont {Dittongo},
  \citenamefont {Grancagnolo}, \citenamefont {Lanceri}, \citenamefont {Vitale},
  \citenamefont {Azzolini}, \citenamefont {Martinez-Vidal}, \citenamefont
  {Panvini}, \citenamefont {Banerjee}, \citenamefont {Bhuyan}, \citenamefont
  {Brown}, \citenamefont {Fortin}, \citenamefont {Hamano}, \citenamefont
  {Kowalewski}, \citenamefont {Nugent}, \citenamefont {Roney}, \citenamefont
  {Sobie}, \citenamefont {Back}, \citenamefont {Harrison}, \citenamefont
  {Latham}, \citenamefont {Mohanty}, \citenamefont {Band}, \citenamefont
  {Chen}, \citenamefont {Cheng}, \citenamefont {Dasu}, \citenamefont {Datta},
  \citenamefont {Eichenbaum}, \citenamefont {Flood}, \citenamefont {Graham},
  \citenamefont {Hollar}, \citenamefont {Johnson}, \citenamefont {Kutter},
  \citenamefont {Li}, \citenamefont {Liu}, \citenamefont {Mellado},
  \citenamefont {Mihalyi}, \citenamefont {Mohapatra}, \citenamefont {Pan},
  \citenamefont {Pierini}, \citenamefont {Prepost}, \citenamefont {Tan},
  \citenamefont {Wu}, \citenamefont {Yu},\ and\ \citenamefont
  {Neal}}]{Aubert2006a}%
  \BibitemOpen
\bibfield  {author} {\bibinfo {author} {\bibfnamefont {B.}~\bibnamefont
  {Aubert}}, \bibnamefont {et~al.},\ }\href {\doibase
  10.1103/PhysRevD.73.052003} {\bibfield  {journal} {\bibinfo  {journal} {Phys.
  Rev. D}\ }\textbf {\bibinfo {volume} {73}},\ \bibinfo {pages} {052003}
  (\bibinfo {year} {2006})}\BibitemShut {NoStop}%
\bibitem [{\citenamefont {Akhmetshin}\ \emph {et~al.}(2013)\citenamefont
  {Akhmetshin}, \citenamefont {Anisenkov}, \citenamefont {Anokhin},
  \citenamefont {Aulchenko}, \citenamefont {Banzarov}, \citenamefont {Barkov},
  \citenamefont {Bashtovoy}, \citenamefont {Berkaev}, \citenamefont {Bondar},
  \citenamefont {Bragin}, \citenamefont {Eidelman}, \citenamefont {Epifanov},
  \citenamefont {Epshteyn}, \citenamefont {Fedotovich}, \citenamefont
  {Gayazov}, \citenamefont {Grebenuk}, \citenamefont {Grigoriev}, \citenamefont
  {Gromov}, \citenamefont {Ignatov}, \citenamefont {Karpov}, \citenamefont
  {Kazanin}, \citenamefont {Khazin}, \citenamefont {Koop}, \citenamefont
  {Kozyrev}, \citenamefont {Krokovny}, \citenamefont {Kuzmenko}, \citenamefont
  {Kuzmin}, \citenamefont {Logashenko}, \citenamefont {Lysenko}, \citenamefont
  {Lukin}, \citenamefont {Mikhailov}, \citenamefont {Pestov}, \citenamefont
  {Perevedentsev}, \citenamefont {Pirogov}, \citenamefont {Pivovarov},
  \citenamefont {Popov}, \citenamefont {Popov}, \citenamefont {Redin},
  \citenamefont {Rogovsky}, \citenamefont {Romanov}, \citenamefont {Ruban},
  \citenamefont {Ryskulov}, \citenamefont {Ryzhenenkov}, \citenamefont
  {Shebalin}, \citenamefont {Shemyakin}, \citenamefont {Shwartz}, \citenamefont
  {Shwartz}, \citenamefont {Sibidanov}, \citenamefont {Shatunov}, \citenamefont
  {Shatunov}, \citenamefont {Snopkov}, \citenamefont {Solodov}, \citenamefont
  {Titov}, \citenamefont {Talyshev}, \citenamefont {Vorobiov}, \citenamefont
  {Yudin},\ and\ \citenamefont {Zaytsev}}]{Akhmetshin2013}%
  \BibitemOpen
\bibfield  {author} {\bibinfo {author} {\bibfnamefont {R.}~\bibnamefont
  {Akhmetshin}}, \bibnamefont {et~al.},\ }\href {\doibase
  10.1016/j.physletb.2013.04.065} {\bibfield  {journal} {\bibinfo  {journal}
  {Phys. Lett. B}\ }\textbf {\bibinfo {volume} {723}},\ \bibinfo {pages} {82}
  (\bibinfo {year} {2013})}\BibitemShut {NoStop}%
\bibitem [{\citenamefont {Lukin}\ \emph {et~al.}(2015)\citenamefont {Lukin},
  \citenamefont {Anisenkov}, \citenamefont {Aulchenko}, \citenamefont
  {Akhmetshin}, \citenamefont {Banzarov}, \citenamefont {Bashtovoy},
  \citenamefont {Berkaev}, \citenamefont {Bragin}, \citenamefont {Vorobiov},
  \citenamefont {Gayazov}, \citenamefont {Grebenuk}, \citenamefont {Grigoriev},
  \citenamefont {Gromov}, \citenamefont {Epifanov}, \citenamefont {Erofeev},
  \citenamefont {Zharinov}, \citenamefont {Ignatov}, \citenamefont {Karpov},
  \citenamefont {Kazanin}, \citenamefont {Kirpotin}, \citenamefont {Koop},
  \citenamefont {Kovalenko}, \citenamefont {Kozyrev}, \citenamefont {Kozyrev},
  \citenamefont {Krokovny}, \citenamefont {Kuzmenko}, \citenamefont {Kuzmin},
  \citenamefont {Logashenko}, \citenamefont {Lysenko}, \citenamefont
  {Mikhailov}, \citenamefont {Okhapkin}, \citenamefont {Pestov}, \citenamefont
  {Perevedentsev}, \citenamefont {Popov}, \citenamefont {Popov}, \citenamefont
  {Razuvaev}, \citenamefont {Rogovsky}, \citenamefont {Romanov}, \citenamefont
  {Ruban}, \citenamefont {Ryskulov}, \citenamefont {Ryzhenenkov}, \citenamefont
  {Sibidanov}, \citenamefont {Solodov}, \citenamefont {Titov}, \citenamefont
  {Talyshev}, \citenamefont {Fedotovich}, \citenamefont {Khazin}, \citenamefont
  {Shatunov}, \citenamefont {Shatunov}, \citenamefont {Shwartz}, \citenamefont
  {Shwartz}, \citenamefont {Shebalin}, \citenamefont {Shemyakin}, \citenamefont
  {Eidelman}, \citenamefont {Epshteyn},\ and\ \citenamefont
  {Yudin}}]{Lukin2015}%
  \BibitemOpen
\bibfield  {author} {\bibinfo {author} {\bibfnamefont {P.~A.}\ \bibnamefont
  {Lukin}}, \bibnamefont {et~al.},\ }\href {\doibase 10.1134/S1063778815020209}
  {\bibfield  {journal} {\bibinfo  {journal} {Phys. At. Nucl.}\ }\textbf
  {\bibinfo {volume} {78}},\ \bibinfo {pages} {353} (\bibinfo {year}
  {2015})}\BibitemShut {NoStop}%
\bibitem [{\citenamefont {Haidenbauer}\ \emph {et~al.}(2015)\citenamefont
  {Haidenbauer}, \citenamefont {Hanhart}, \citenamefont {Kang},\ and\
  \citenamefont {Mei{\ss}ner}}]{Haidenbauer2015}%
  \BibitemOpen
\bibfield  {author} {\bibinfo {author} {\bibfnamefont {J.}~\bibnamefont
  {Haidenbauer}}, \bibnamefont {et~al.},\ }\href {\doibase
  10.1103/PhysRevD.92.054032} {\bibfield  {journal} {\bibinfo  {journal} {Phys.
  Rev. D}\ }\textbf {\bibinfo {volume} {92}},\ \bibinfo {pages} {054032}
  (\bibinfo {year} {2015})}\BibitemShut {NoStop}%
\bibitem [{\citenamefont {Berestetskii}\ \emph {et~al.}(1982)\citenamefont
  {Berestetskii}, \citenamefont {Lifshitz},\ and\ \citenamefont
  {Pitaevskii}}]{landau4}%
  \BibitemOpen
\bibfield  {author} {\bibinfo {author} {\bibfnamefont {V.~B.}\ \bibnamefont
  {Berestetskii}}, \bibinfo {author} {\bibfnamefont {E.~M.}\ \bibnamefont
  {Lifshitz}},\ \bibnamefont {and}\ \bibinfo {author} {\bibfnamefont {L.~P.}\
  \bibnamefont {Pitaevskii}},\ }\href@noop {} {\emph {\bibinfo {title}
  {{Quantum Electrodynamics}}}},\ \bibinfo {edition} {2nd}\ ed.,\ \bibinfo
  {publisher} {Pergamon Press}, \bibinfo {address} {Oxford}\BibitemShut
  {NoStop}%
\bibitem [{\citenamefont {Fadin}\ and\ \citenamefont
  {Khoze}(1987)}]{Fadin1987}%
  \BibitemOpen
\bibfield  {author} {\bibinfo {author} {\bibfnamefont {V.~S.}\ \bibnamefont
  {Fadin}}\ \bibnamefont {and}\ \bibinfo {author} {\bibfnamefont {V.~A.}\
  \bibnamefont {Khoze}},\ }\href
  {http://www.jetpletters.ac.ru/ps/1234/article_18631.pdf
  http://inspirehep.net/record/251362?ln=en} {\bibfield  {journal} {\bibinfo
  {journal} {JETP Lett.}\ }\textbf {\bibinfo {volume} {46}},\ \bibinfo {pages}
  {525} (\bibinfo {year} {1987})}\BibitemShut {NoStop}%
\bibitem [{\citenamefont {Ericson}\ and\ \citenamefont
  {Weise}(1988)}]{Ericson1988}%
  \BibitemOpen
\bibfield  {author} {\bibinfo {author} {\bibfnamefont {T.~E.~O.}\ \bibnamefont
  {Ericson}}\ \bibnamefont {and}\ \bibinfo {author} {\bibfnamefont
  {W.}~\bibnamefont {Weise}},\ }\href
  {https://books.google.ru/books/about/Pions_and_nuclei.html?id=v099AAAAIAAJ&pgis=1}
  {\emph {\bibinfo {title} {{Pions and nuclei}}}},\ \bibinfo  {publisher}
  {Clarendon Press}, \bibinfo {address} {Oxford}\BibitemShut {NoStop}%
\bibitem [{\citenamefont {Badalyan}\ \emph {et~al.}(1982)\citenamefont
  {Badalyan}, \citenamefont {Kok}, \citenamefont {Polikarpov},\ and\
  \citenamefont {Simonov}}]{Badalyan1982}%
  \BibitemOpen
\bibfield  {author} {\bibinfo {author} {\bibfnamefont {A.}~\bibnamefont
  {Badalyan}}, \bibnamefont {et~al.},\ }\href {\doibase
  10.1016/0370-1573(82)90014-X} {\bibfield  {journal} {\bibinfo  {journal}
  {Phys. Rep.}\ }\textbf {\bibinfo {volume} {82}},\ \bibinfo {pages} {31}
  (\bibinfo {year} {1982})}\BibitemShut {NoStop}%
\bibitem [{\citenamefont {Amsler}\ \emph {et~al.}(2003)\citenamefont {Amsler},
  \citenamefont {Anisovich}, \citenamefont {Baker}, \citenamefont {Barnett},
  \citenamefont {Batty}, \citenamefont {Benayoun}, \citenamefont {Bl{\"{u}}m},
  \citenamefont {Braune}, \citenamefont {Bugg}, \citenamefont {Case},
  \citenamefont {Cred{\'{e}}}, \citenamefont {Crowe}, \citenamefont {Doser},
  \citenamefont {D{\"{u}}nnweber}, \citenamefont {Engelhardt}, \citenamefont
  {Faessler}, \citenamefont {Haddock}, \citenamefont {Heinsius}, \citenamefont
  {Heinzelmann}, \citenamefont {Hessey}, \citenamefont {Hidas}, \citenamefont
  {Jamnik}, \citenamefont {Kalinowsky}, \citenamefont {Kammel}, \citenamefont
  {Kisiel}, \citenamefont {Klempt}, \citenamefont {Koch}, \citenamefont
  {Kunze}, \citenamefont {Kurilla}, \citenamefont {Landua}, \citenamefont
  {Matth{\"{a}}y}, \citenamefont {Meyer}, \citenamefont {Meyer-Wildhagen},
  \citenamefont {Montanet}, \citenamefont {Ouared}, \citenamefont {Peters},
  \citenamefont {Pick}, \citenamefont {Popkov}, \citenamefont {Ratajczak},
  \citenamefont {Regenfus}, \citenamefont {Reinnarth}, \citenamefont {Roethel},
  \citenamefont {Sarantsev}, \citenamefont {Spanier}, \citenamefont
  {Strohbusch}, \citenamefont {Suffert}, \citenamefont {Suh}, \citenamefont
  {Thoma}, \citenamefont {Uman}, \citenamefont {Wallis-Plachner}, \citenamefont
  {Walther}, \citenamefont {Wiedner}, \citenamefont {Wittmack},\ and\
  \citenamefont {Zou}}]{Amsler2003}%
  \BibitemOpen
\bibfield  {author} {\bibinfo {author} {\bibfnamefont {C.}~\bibnamefont
  {Amsler}}, \bibnamefont {et~al.},\ }\href {\doibase
  10.1016/S0375-9474(03)00912-6} {\bibfield  {journal} {\bibinfo  {journal}
  {Nucl. Phys. A}\ }\textbf {\bibinfo {volume} {720}},\ \bibinfo {pages} {357}
  (\bibinfo {year} {2003})}\BibitemShut {NoStop}%
\bibitem [{\citenamefont {Klempt}\ \emph {et~al.}(2005)\citenamefont {Klempt},
  \citenamefont {Batty},\ and\ \citenamefont {Richard}}]{Klempt2005}%
  \BibitemOpen
\bibfield  {author} {\bibinfo {author} {\bibfnamefont {E.}~\bibnamefont
  {Klempt}}, \bibinfo {author} {\bibfnamefont {C.}~\bibnamefont {Batty}},\
  \bibnamefont {and}\ \bibinfo {author} {\bibfnamefont {J.~M.}\ \bibnamefont
  {Richard}},\ }\href {\doibase 10.1016/j.physrep.2005.03.002} {\bibfield
  {journal} {\bibinfo  {journal} {Phys. Rep.}\ }\textbf {\bibinfo {volume}
  {413}},\ \bibinfo {pages} {197} (\bibinfo {year} {2005})}\BibitemShut
  {NoStop}%
\end{thebibliography}
\end{document}